\documentclass[11pt]{article}
\usepackage{fullpage}
\usepackage[margin=1in]{geometry}
\def\anonymous{0}
\def\comments{0}

\usepackage[english]{babel}
\usepackage{hyperref}
\hypersetup{
  colorlinks   = true, %
  urlcolor     = blue, %
  linkcolor    = blue,  %
  citecolor   = red    %
}
\usepackage{url}

\input{./notation.sty}

\usepackage[textwidth=2.5cm]{todonotes}

\newcommand{\mnote}[3]{\todo[color=#3!40,size=\footnotesize]{\textbf{#2:} #1}}

\ifnum\comments=0
\newcommand{\anote}[1]{}
\newcommand{\bnote}[1]{}
\newcommand{\pnote}[1]{}
\newcommand{\tnote}[1]{}
\newcommand{\snote}[1]{}
\else
\newcommand{\anote}[1]{\mnote{#1}{A}{white}}
\newcommand{\bnote}[1]{\mnote{#1}{B}{cyan}}
\newcommand{\pnote}[1]{\mnote{#1}{P}{blue}}
\newcommand{\tnote}[1]{\mnote{#1}{T}{green}}
\newcommand{\snote}[1]{\mnote{#1}{S}{gray}}
\fi

\newcommand{\theExponent}{\Omega(1)}

\title{Derandomized Sunflowers and Radical Lower Bounds}
\ifnum\anonymous=1
\author{}
\date{}
\else
\author{Bruno Cavalar
\thanks{Department of Computer Science, University of Oxford, United Kingdom.
Supported by the UK Research and Innovation (UKRI) via the EPSRC project EP/Z534158/1 on ``Integrated Approach to Computational Complexity: Structure, Self-Reference and Lower Bounds'' and through an EPSRC Postdoctoral Fellowship (grant ref.\! UKRI3441).
Email: \texttt{bruno.cavalar@cs.ox.ac.uk}}
\and Th\'{e}o Bor\'{e}m Fabris\thanks{Department of Computer Science,
University of Copenhagen, Denmark. Supported by the European Research
Council (ERC) under grant agreement no. 101125652 (ALBA).
The author thanks BARC (VILLUM grant no. 54451) for the supportive research environment.
Email:
\texttt{thfa@di.ku.dk}}
\and Partha Mukhopadhyay\thanks{Chennai Mathematical Institute, India.
Email: \texttt{partham@cmi.ac.in}}
\and Srikanth Srinivasan\thanks{Department of Computer Science, University
of Copenhagen, Denmark. Supported by the European Research Council (ERC)
under grant agreement no. 101125652 (ALBA). Email: \texttt{srsr@di.ku.dk} }
\and Amir Yehudayoff\thanks{Department of Computer Science, The University
of Copenhagen, and The Technion-IIT. Supported by DNRF Chair grant no. 20231101-28697.
Email: \texttt{amir.yehudayoff@gmail.com}}
}
\fi

\begin{document}

\maketitle

\vspace{-1em}
\begin{abstract}
Robust Sunflower lemmas imply that any large enough monotone DNF of width $w$ contains a sunflower, i.e., a DNF equivalent to the conjunction of a common core with a DNF that is heavily biased towards $1$. While these lemmas are typically proved in the context of the uniform distribution or i.i.d. distributions, they also hold for $O(w^2)$-wise independent distributions by Bazzi's theorem (Bazzi (SICOMP 2009); Tal (CCC 2017)), which states that width-$w$ DNFs cannot distinguish between such distributions and the uniform distribution.

We prove a derandomization of robust sunflower lemmas for $O(w)$-wise independent distributions, which recovers the derandomization above with slightly weaker parameters. Notably, our result also holds for distributions whose marginals are ``locally independent'' and ``approximately identically'' distributed. It is not known whether such distributions fool general DNFs, which makes the derandomization via DNF-fooling techniques unavailable. Our proof of this derandomization is via the method of moments, which might be of independent interest.

We apply this result to obtain improved separations between non-monotone arithmetic circuits and the radical of monotone circuits. Specifically, we construct an \emph{explicit} family of multilinear polynomials $P_n$ that have small non-monotone arithmetic circuits such that any power of $P_n$ cannot be computed by monotone arithmetic circuits of \emph{sub-exponential} size. This improves on a recent result of Cavalar, Fabris, Mukhopadhyay, Srinivasan and Yehudayoff (STOC 2026), who proved a non-explicit separation that was quasipolynomial. Our construction uses Nisan-Wigderson designs with an additional linear-algebraic expansion property.
\end{abstract}

\thispagestyle{empty} %

\newpage
\tableofcontents

\thispagestyle{empty} %
\newpage

\setcounter{page}{1} %
\section{Introduction}
\label{sectIntro}

\subsection{Background and Results}

The sunflower lemma of 
Erd\H{o}s and Rado \cite{ER60}
states that every large enough 
family of sets must contain a \emph{sunflower}:
a family of sets whose pairwise intersections are identical.
Along the decades, this 
elegant 
definition 
has found applications in several areas of theoretical computer 
science,
including
circuit complexity~\cite{Raz85,DBLP:journals/cc/HastadJP95},
data structure lower bounds~\cite{FrandsenMiltersenSkyum1993,DBLP:conf/coco/RamamoorthyR18},
matrix multiplication algorithms~\cite{DBLP:journals/cc/AlonSU13},
parametrized complexity~\cite{DBLP:series/txtcs/FlumG06}, 
and much more~\cite{Rao2025}.

An ``approximate'' version of sunflowers was defined by~\cite{Rossman2014}
in the context of average-case monotone circuit complexity.
This variant, nowadays called \emph{robust sunflowers},
has since become a central key in several results and breakthroughs,
including progress in the sunflower conjecture~\cite{ALWZ21},
improved monotone circuit lower bounds~\cite{CKR22,deRezendeVinyals25,CGRSS25,DBLP:conf/coco/BlasiokM25,CFMSY25},
communication complexity~\cite{DBLP:conf/innovations/LovettMMPZ22},
DNF sparsification~\cite{DBLP:journals/cc/GopalanMR13},
sampling lower bounds~\cite{DBLP:conf/approx/FilmusLR023},
top-down circuit lower bounds~\cite{DBLP:conf/focs/GoosRS023},
etc.

A robust sunflower can be defined as a monotone DNF
which accepts
inputs randomly sampled from a product distribution
with very high probability,
provided that the literals present in all terms, i.e., the core of the sunflower,
are set to 1.
Rossman~\cite{Rossman2014} proved the first bounds
on how large a DNF must be to contain a robust sunflower as a sub-DNF,
and later
optimal bounds were discovered
by~\cite{ALWZ21,Rao2020,BellChueluechaWarnke2021}.

There is nothing special about product distributions, and one could also
define robust sunflowers with respect to other distributions,
particularly distributions with significantly smaller entropy, which we call derandomized sunflowers.
Proving a robust sunflower lemma over such distributions
was the key behind the recent exponential monotone circuit lower bound for
Perfect-Matching~\cite{CGRSS25},
as well as an improved monotone circuit lower bound for $\NC$~\cite{CFMSY25}.
It has also been speculated that such ``derandomized sunflowers'' 
may provide a path towards improved lifting theorems in communication
complexity~\cite{Kelley21}.

\paragraph{Robust sunflower bounds for almost-independent distributions.}

In~\cite{CGRSS25}, 
a derandomized sunflower lemma
was obtained by
an ad-hoc reduction to fully random robust sunflowers.
The reduction works only for DNFs whose terms form a partial matching
and to a specific distribution,
being thus more tailored to their specific application.
In~\cite{CFMSY25}, 
a derandomized sunflower lemma
was instead obtained 
by 
employing
Bazzi's
theorem~\cite{Bazzi09,Tal17}
to
argue that small DNFs cannot distinguish between
$k$-wise independent and fully random distributions.
This argument works for all small DNFs and $k$-wise independent distributions.

Unfortunately, a significant downside of 
employing pseudorandom generators (PRGs) for DNFs,
\tnote{Review 1.2 complained about our use of the term PRGs, and I partially agree that we are not really using PRGs but just fooling arguments}
such as Bazzi's theorem,
as a blackbox
is that
such reductions are potentially suboptimal.
A standard structure-versus-randomness argument implies that
to obtain robust sunflower bounds
it suffices to show that spread DNFs are robust sunflowers,
where we say that a DNF is \emph{spread} if no set of literals appears in 
disproportionally too many terms.
Bazzi's theorem, however, applies to \emph{all} DNFs of sufficiently small
size, not merely the spread ones.
This generality comes with a cost:
the resulting sunflower lemma has significantly worse bounds
than the fully random case.
In particular, the authors of~\cite{CFMSY25} only manage
to obtain 
exponential lower bounds with exponent $n^{1/3}$ on the monotone complexity
of $\NC$ functions, whereas
exponential lower bounds with exponent 
$n^{1/2}$
are the state-of-the-art for $\NP$ functions~\cite{CKR22}.
More fundamentally, 
such arguments cannot capture distributions that do not fool DNFs.
As an example, almost-independent distributions cannot even fool
DNFs of polylogarithmic width~\cite{DBLP:conf/coco/AlekseevGGMR0025}.\tnote{Review 2.3 complained about this claim.}

Our first contribution is to show robust sunflower bounds
for ``local'' convex combinations of product distributions via a high-order Markov's inequality,
which avoids the more sophisticated machinery from PRGs for DNFs.
We say that a distribution $\cald$ with support $\blt^n$ 
is
\emph{$(k,\eps)$-pseudo-independent with marginal probability $q$}
if, for every set $S \sseq [n]$ of size at most $k$,
the marginal distribution $\cald_S$ of $\cald$ on $\blt^S$
is a convex combination of product distributions,
in each of which every bit is $1$ with probability 
$q' \in [(1-\eps)q, (1+\eps)q]$.
A monotone DNF $\calf$ is a 
\emph{$(\cald,\delta)$-sunflower}
if
$\Exp_{x \flws \cald}[\calf_\Delta(x)] > 1-\delta$,
where $\calf_\Delta$
is the DNF obtained from $\calf$ by setting all
literals that appear in
\emph{all the terms} of $\calf$ to 1.
We prove the following:

\begin{theorem}
    \label{thm:sf-intro}
    For every $w\in\Naturals$ and $0 < q,\delta < 1$,
    let $\calf$ be a width-$w$ monotone DNF 
    of size at least
    $(\frac{w}{q}\log(1/\delta))^w$.
    There exist parameters $k = O(w\log(1/\delta))$ and
    $\varepsilon = \Omega(1/w)$
    such that
    if $\cald$ is $(k,\eps)$-pseudo-independent with marginal distribution $q$,
    then $\calf$ contains a $(\cald,\delta)$-sunflower.
\end{theorem}

Let us compare the above result with the sunflower lemma
of~\cite{CFMSY25}.
Their arguments
roughly show 
a sunflower bound of $(\log(1/\delta)/q)^w$
for parameters $k \approx w^2 + w\log(1/\delta)$
and $\eps = 0$,
using the best known bounds for $k$-wise independence needed to fool DNFs~\cite{Tal17}.
Whereas their size bound is better for this regime of parameters,
our result achieves a derandomization of sunflowers with smaller $k$
even in the $\eps = 0$ regime, as we only need $k = O(w \log (1/\delta))$.
Moreover, 
it is easy to check that 
$(k,\eps)$-pseudo-independent distributions with sufficiently small
$\eps$ are also ``\emph{almost independent}''\footnote{
    A distribution 
    $\cald$ is
    \emph{$(k,\eps)$-wise almost independent} 
    if $(1-\eps)2^{-k} \leq \Pr_{x \flws \cald}[x_A = \alpha] \leq
    (1+\eps)2^{-k}$ for every $\alpha \in \blt^{A}$
    and every $A$ of size at most $k$.
};
whereas some ``\emph{almost independent}'' distributions
cannot fool DNFs~\cite{DBLP:conf/coco/AlekseevGGMR0025}.
Our result thus suggests that derandomizing sunflower lemmas 
may be strictly easier than
building PRGs for DNFs.\tnote{Again there was a complain about our use of PRGs: Reviews 1.2 and 2.2}

Given the wide applicability and consequences of robust sunflower bounds,
we expect our derandomized sunflower bounds to be useful in other contexts, and
our techniques also seem sufficiently flexible to be
adaptable to different distributions, as we discuss in
\Cref{sec:techniques}.
We now describe a particular application we derive from our derandomized
sunflower lemma, solving an open problem in circuit complexity~\cite{CFMSY25}.
The \emph{pseudo-independence} condition (and not just $k$-wise independence)
allowed by Theorem~\ref{thm:sf-intro} is needed for this application.

\paragraph{Monotone radical lower bounds.}
Monotone arithmetic complexity studies arithmetic circuits over the nonnegative reals that disallow subtraction.
The study of monotone polynomials is natural for counting problems, since many counting questions about combinatorial structures are naturally characterized as evaluating polynomials with non-negative coefficients on Boolean inputs.
A classical example is the problem of counting the number of perfect matchings in bipartite graphs and its characterization via evaluations the permanent polynomial. %
Furthermore,
many natural combinatorial strategies for solving these problems (e.g., pure dynamic programming algorithms) can be phrased as monotone arithmetic circuits~\cite{JS20},
and algorithms that avoid subtractions and cancellations are inherently more robust to numerical noise, e.g. from rounding errors~\cite{Demmel,Fomin}.

This led researchers to the question of understanding the gap in computational power between monotone and non-monotone computations and by now, many results in this direction are known~\cite{ShamirSnir1977,Valiant80, JerrumSnir1982, GashkovSergeev1991, RazYehudayoff11, CDM21, CDGM22, CFMSY25}. %
The seminal work of Valiant~\cite{Valiant80} showed that a nonmonotone circuit computing a monotone polynomial using only one negation gate can be exponentially more powerful than any monotone computation.
Continuing this line of work, the current state-of-the-art results establish strongly exponential separations between depth-three non-monotone circuits and general monotone circuits~\cite{CFMSY25}, showing the best possible exponential separation between nonmonotone and monotone classes.
These results show that monotone counting algorithms are inherently weaker than their non-monotone counterparts.

However, it may be possible to get around this weakness of monotone computation with a trick:
first perform an efficient monotone computation $M$ that produces a function $g(f)$ of the answer $f$ and
then invert the function $g$ using a simple non-monotone computation $C$ to produce $f$ itself.
In other words, $f$ can be computed by the composition $C\after M$ of $C$ after $M$,
which can be interpreted as way of computing $f$ via a non-monotone circuit with restricted non-monotone computation $C$.
\tnote{Added an explanation about motivation for the radical class.}

A natural option for the non-monotone computation $C$ is taking the $k$-th root of $M$, since there are fast and numerically robust algorithms for computing $k$-th roots as long as $k$ itself is not too large~\cite{Kaltofen87}.
The resulting question can then be naturally stated in the language of \emph{radicals}. 
Radicals appear, for example, in geometry (via the Pythagorean theorem) and in solving quadratic equations. 
More abstractly,
a radical of an ideal is the set of elements whose powers belong to the ideal. 
There are many examples of radicals that are larger than their ideals,
such as the radical of the ideal generated by $2$ in the ring of algebraic integers and the radical of the ideal generated by $x^2$ in the ring of polynomials.
Another important example is Hilbert's Nullstellensatz, which characterizes the geometric consequences of a set of polynomial equations in terms of its (algebraically defined) radical.
We define the \emph{radical $\sqrt{\calc}$ of a circuit class $\calc$} as the set of polynomials $P$ such that some power of $P$ is in $\calc$. 
For non-monotone classes $\Ccal$, there are questions about $\sqrt{\Ccal}$ that are relevant to the study of Hardness vs. Randomness in arithmetic circuits~\cite{Andrews}.

It is then natural to investigate potential computational gaps between non-monotone computation and monotone radical computations.
This question can be formalized as follows: is there a polynomial $P(x_1,\ldots, x_n)$ that has non-monotone circuits of size $\poly(n)$, but is not in the radical $\sqrt{\calc}$,\footnote{We could also bound the order of the resulting root to a polynomial function of $n$. But we keep this stronger definition as our results apply even in this setting.} where 
$\calc$ is the class of \emph{monotone} arithmetic circuits of some bounded size $s(n)$? 
Most existing proof techniques in monotone arithmetic complexity --- either support-based methods \cite{ShamirSnir1977,Valiant80, JerrumSnir1982, GashkovSergeev1991} or approaches based on communication complexity and related techniques~\cite{RazYehudayoff11, Yehudayoff19, Srinivasan19, CDM21} --- appear to fall short of resolving the question about the relative power of nonmonotone classes and the radical of monotone circuits.

Very recently, the work of~\cite{CFMSY25} made progress on this problem, by showing the first separation between $\VP$ and the radical of monotone circuits with $s(n)$ being a superpolynomial function of $n$.
Their result was obtained via a connection to monotone \emph{Boolean} circuit lower bounds, suggested by Hrube\v{s} and Yehudayoff~\cite{HY21} in the context of proving lower bounds for monotone circuits with division gates.
For every $n$-variate monotone polynomial $P$ over $\mathbb{R}$, define a monotone Boolean function $f_P : \{0,1\}^{n} \to \{0,1\}$ via $f_P(x) = 1$ iff $P(x) > 0$. The key observation is that a monotone arithmetic circuit for $P$ naturally yields a monotone Boolean circuit for $f_P$ by replacing $+$ gates with $\lor$ gates, and $\times$ gates with $\land$ gates, and positive $\Reals$-constant with the bit $1$. Importantly, the function $f_P$ depends only on the set of points where $P$ evaluates to a non-zero value and is therefore invariant under taking powers of $P$. Consequently, a monotone Boolean lower bound for $f_P$ implies monotone arithmetic lower bounds for all powers of $P$. 

However, the polynomials constructed in~\cite{CFMSY25} are not explicit, as their definitions rely on a base matrix chosen probabilistically, so their circuit upper bound (in $\VP$) is not known to be described by a Turing machine running in polynomial time.
\tnote{Review 3.2 complained about notions of explicit vs. fully explicit}
Moreover, their monotone lower bound is only quasi-polynomial, whereas previous monotone vs. non-monotone separations (which do not imply lower bounds for the radical) yield exponential bounds. 

We remove both these shortcomings.
Using ideas from pseudorandomness
--- more precisely, by understanding linear-algebraic expansion properties of Nisan-Wigderson designs~\cite{NisanWigderson1994} based on univariate polynomials ---
we explicitly construct polynomials $(P_m)_m$ computed by arithmetic circuits described by Turing machine running in deterministic polynomial time and having the properties needed to obtain an exponential radical monotone lower bound
via our derandomized sunflower lemma~(\Cref{thm:sf-intro}).
Our lower bound is obtained via
a criterion of~\cite{CFMSY25} that reduces monotone Boolean
circuit lower bounds to sunflower bounds over carefully chosen pseudo-independent distributions.

\begin{theorem}
    \label{thm:radical-sep-intro}
    There is an infinite sequence $(P_m(x_1,\ldots,x_m))_{m\in\Naturals}$ of  polynomials
    such that the following hold:
    \begin{enumerate}
        \item (Explicit Upper bound) $P_m$ is a monotone multilinear polynomial with an arithmetic circuit $C_m$ of size $m^{O(1)}$. Further, given $m$, the circuit $C_m$ can be computed in deterministic time $m^{O(1)}$.
        \item (Radical Lower bound) Any monotone arithmetic circuit computing a power of $P_m$ must have size at least $\exp(m^{\theExponent}).$
    \end{enumerate}
\end{theorem}

Employing arithmetic depth-reduction results~\cite{AgrawalVinay08, Koiran12, Tavenas15, GKKS16},
we also obtain the first super-polynomial
separation between non-monotone constant-depth circuits and the radical of monotone circuits.

\begin{corollary}
    \label{cor:radical-ac0-intro}
    For every constant $k \geq 1$,
    there is an infinite sequence $(Q_M(x_1,\ldots,x_M))_{M\in\Naturals}$
    of polynomials
    such that the following hold:
    \begin{enumerate}
        \item $Q_M$ is a monotone multilinear polynomial with a depth $O(k)$ arithmetic circuit $C'_M$ of size $M^{O(1)}$. Further, given $M$, the circuit $C'_M$ can be computed in deterministic time $M^{O(1)}$.
        \item Any monotone arithmetic circuit computing a power of $Q_M$ must have size at least
        $M^{\Omega((\log M)^k)}$.
    \end{enumerate}
\end{corollary}

\subsection{Challenges and Proof Ideas}
\label{sec:techniques}

We now describe the main challenges and proof ideas in our results.
First, we describe how to prove \Cref{thm:sf-intro}, our derandomized sunflower lemma.
Then we show how the sunflower lemma can be employed to obtain the
separation of \Cref{thm:radical-sep-intro} for polynomials with certain
properties.
We then describe how to construct such polynomials.

\paragraph{Derandomization of sunflowers via the method of moments.}

The first proof of a robust sunflower lemma was given by
Rossman~\cite{Rossman2014} by means of an application of
Janson's inequality~\cite{https://doi.org/10.1002/rsa.3240010209}
(see also~\cite[Chapter 8]{AlonSpencerTPM16}).
Essentially, Janson's inequality helps us analyse the following 
type of event.
Let $\calf$ be a width-$w$ monotone
DNF over $n$ bits with terms $F_1, \dots, F_N$
where $N = r^w$ for some parameter $r$.
We identify a term with its set of literals, thus frequently referring to $F_i$ both as a conjunction and as a set.
Let $x$ be a uniformly random string of $n$ bits.
Write $i \sim j$ if the events 
$\set{F_i(x)=1}$ and $\set{F_j(x)=1}$ 
are dependent 
(i.e., if $F_i \cap F_j \neq \emptyset$).
For $\mu \defeq \sum_{i\in [N]} \Exp_{x \flws \blt^n}[F_i(x)]$,
Janson's inequality gives the following bound:
\begin{equation*}
    \Pr_{x \flws \blt^n}[\calf(x) = 0]
    \leq
    \exp(-\mu^2/(2\Delta)),
\end{equation*}
where $\Delta := \sum_{i \sim j}\Pr_{x \flws \blt^n}[(F_i \land F_j)(x) = 1]$.

This bound is quite useful when $\calf$ is \emph{$r$-spread}, that is,
when 
\[\card{\setst{i \in [N]}{A \sseq F_i}} 
\leq r^{w-\card{A}}\]
for every set $A \sseq [n]$.
One can then upper bound $\Delta$ simply by counting,
for every $F \in \calf$ and $A \sseq F$, 
how many sets in $\calf$  intersect $F$ at $A$;
there will be at most $r^{w-\card{A}}$ such sets.
From here, it follows that spread DNFs are robust sunflowers with an empty
core.

When the DNF is not spread, it suffices to find a maximally spreadness
violating set $A$; we can easily argue that
$\calf_A = \set{F_i \sm A : i\in [N], A \sseq F_i}$ is itself $r$-spread, and
thus $\set{F_i : i\in [N], A \sseq F_i}$ is a robust sunflower inside
$\calf$.
Thus, the technical core of the proof is showing that spread DNFs are
robust sunflowers with respect to the uniform distribution.

This simple argument, despite being suboptimal, still appears in some form
in the optimal robust sunflower bounds obtained by~\cite{ALWZ21,Rao2020},
and thus generalising it 
can be the first step towards
sunflowers bounds for other distributions.
The problem is that Janson's inequality does not necessarily apply for
arbitrary distributions.

Even if the distribution is $t$-wise independent and
there are not many pairwise dependencies between terms (so $\Delta$ is small),
arbitrary correlations can
appear between larger collections of terms.
However, existing proofs of Janson's inequality seem to require
the fact that, if 
a random variable $Y_{i} := \set{F_i(x) = 1}$, for $i \in [N]$,
is independent of each of the $p$ random variables
$Y_{i_1}(x),\dots,Y_{i_p}(x)$
individually,
then 
$Y_i$
is independent of all of them. This holds for product
distributions
but not necessarily for
a $(t,\eps)$-pseudo-independent distribution $\cald$. 
However, if $p \leq t/w$, the total amount of bits involved is at most
$t$, and we can hope to control the dependencies in this case.

The approach taken by~\cite{CFMSY25} is just to find an Erd\H{o}s-Rado sunflower 
with $p \leq t/w$ petals. This works: it gives a robust sunflower bound, but
the bound is too coarse and only yields quasi-polynomial circuit lower bounds.

Our approach is to apply the \emph{method of moments}.
Letting $\alpha := q^{-w}$ 
(assuming $\cald$ has marginal probability $q$),
we define $X_i := Y_i - \alpha$ and $X \defeq \sum X_i$.
We also let
$\mu \defeq N\alpha$. 
The method of moments is the following observation
from Markov's inequality:
\begin{equation*}
    \Pr_{x \flws \cald}[\calf(x) = 0] = \Pr[X = - \mu]
    \leq
    \Pr[X^p \geq \mu^p]
    \leq
    \frac{\Exp[X^p]}{\mu^p},
\end{equation*}
where $p \leq t/w$ is an even number.
The technical heart is then to bound $\Exp[X^p]$.
We will obtain such an estimate in two steps.
\begin{enumerate}
    \item 
        For any $s \in [p]$, and $i\in[N]^s$, and $X_{i_1},\dots,X_{i_s}$
        whose dependency graph forms a single connected component,
        we give a generic bound on 
        $\abs{\Exp[X_{i_1} \cdot \dots \cdot X_{i_s}]}$
        depending solely on the ``intersection pattern'' of the literals
        in the terms $F_{i_1},\dots,F_{i_s}$.
        Here we use the pseudo-independence of $\cald$.
   \item 
       We partition the
       summands
       $\Exp[X_{i_1} \cdot \dots \cdot X_{i_p}]$
       using the partition of $[p]$ induced by the
       dependency graph 
       corresponding to the terms $F_{i_1},\ldots, F_{i_p}$.  
       We use the $r$-spreadness of $\calf$ to bound the number of 
       summands
       that respect a certain fixed intersection pattern,
       along with a careful analysis of the number of partitions
       that give rise to a specific dependency graph.
\end{enumerate}

The first step avoids the difficulty of reproving Janson's inequality
by focusing only on the expectation of random variables involving at most
$t$ bits of $\cald$. The second step applies a new ``encoding-type''
argument somewhat
reminiscent of some encoding arguments
employed in~\cite{ALWZ21,Rao2020}, albeit in a ``higher-moment'' fashion.
Let
$F_{i_1},\dots,F_{i_p}$ be terms of $\calf$
and 
define
$\Vars(F_{i_j})$ 
and
$\Vars(F_{i_1,\dots,i_{j-1}})$
as the literals involved in $F_{i_j}$
and $F_{i_1},\dots,F_{i_{j-1}}$, respectively.
If we know that 
$\abs{\Vars(F_{i_j}) \cap \Vars(F_{i_1,\dots,i_{j-1}})} = \ell_j > 0$
for every $\ell \in \set{2,\dots,p}$,
we can ``encode'' $F_{i_j}$
by first choosing a set $V$ of size $\ell_j$
inside $\Vars(F_{i_1,\dots,i_{j-1}})$
(there are $\binom{(j-1)w}{\ell_j}$ such sets)
and then choosing $F_{i_j}$
among the $N r^{-\ell_j}$ sets containing $V$.

This argument could form the basis for a more sophisticated proof
giving optimal derandomized sunflower bounds.
Moreover,
our careful partitioning of tuples
$(F_{i_1},\dots,F_{i_p})$ according to their dependencies
(and the computation of the number of such partitionings)
seems highly flexible and potentially adaptable to other families of distributions
which are not pseudo-independent.

\paragraph{Rank-based monotone polynomial.}
Recall that we want to show a monotone Boolean circuit lower bound for
a function of the form $f_P : \blt^m \to \blt$, where
$P$ is a monotone polynomial 
and
$f_P(x) = 1 \iff P(x) > 0$.
Given an infinite sequence of matrices $(M_m)$ where $M_m \in
\mathbb{R}^{n \times m}$ with $n \leq m$ (here $n$ is a parameter that
grows with $m$), we define a sequence of multilinear polynomials with
non-negative real coefficients as follows. For each $m,$ 
\[ P_m(x_1,\ldots, x_m) \defeq \sum_{S\in \binom{[m]}{n}} \det(M_m[S])^2\cdot \prod_{i\in S}x_i 
\] 
where $M_m[S]$ denotes the $n\times n$ submatrix obtained by keeping only
the columns indexed by $S$ (in increasing order). 
Using Cauchy-Binet formula,
it can be observed that the polynomial
family can be computed by a uniform family of circuits if the matrix
sequence is explicit~\cite{CFMSY25}. 
 
Note that the function $f := f_{P_m}$
satisfies $f(1_S) = 1 \iff M_m[S]$ is full-rank,
where $1_S$ is the indicating vector of a set columns $S \sseq [m]$.
For simplicity, henceforth we denote $M_m$ by $M$.

\paragraph{Exponential monotone lower bounds via the approximation method.}
To show a lower bound for $f$,
we employ a monotone circuit lower bound criterion from~\cite{CFMSY25}
that reduces monotone circuit lower bounds to robust sunflower bounds.
The criterion states that it suffices to define a no-distribution $\cald_0$ and
a yes-distribution $\cald_1$ over the inputs $\bcube{m}$ with the following
properties:
\begin{enumerate}
    \item 
        The distribution $\cald_1$ must be 
        \emph{$Q$-spread}:
        \tnote{Review 3.12 complained about this name, and suggested to call it $Q$-bounded.}
        $\Pr_{x \flws \cald_1}[\bigwedge_{i \in A} x_i = 1]
        \leq Q^{-\card{A}}$,
        for every $A \sseq [m]$;
    \item 
        Any monotone DNF with at least $r^\ell$ terms of width 
        $\ell \in [t]$
        must contain a $(\cald_0,m^{-3t})$-sunflower; and
    \item $\Pr_{b\sim \blt, x \sim \cald_b}[f(x)=b] = \Omega(1)$.
\end{enumerate}
This results in a monotone circuit lower bound 
which is
roughly
$(Q/r)^{\Omega(t)}$.

In order to construct distributions $\cald_1$ and $\cald_0$ satisfying the above
properties, 
we will 
need 
the columns of the matrix $M$ to have 
the following properties (informally described):
\begin{enumerate}
    \item 
        \emph{Linear-algebraic expansion}: given any
        subspace $U$ spanned by a sub-collection of the columns, the chance
        that a uniformly chosen column vector will be outside of $U$ is
        high; 
    \item \emph{Design property:}
        any pair of columns should intersect only at a few locations in their supports.
\end{enumerate}

Property (1) allows us to show that a random set of $\approx n\log n$
columns from $M$ is very likely to be full-rank: indeed, if we imagine
we are sampling those columns one-by-one, the expansion property makes it
unlikely that the next column is spanned by the previous columns.
We thus get a distribution $\cald_1$ of accepting inputs which is
roughly 
$(m/(n\log n))$-spread
as it is the behaviour of the distribution obtained by uniformly choosing $n\log n$ columns of $M$.

Property (2) allows us to show a robust sunflower bound for the following
distribution $\cald_0$:
sample $u \in \set{-1,+1}^n$ uniformly at random,
and let $x_i \defeq 1$ if the inner-product $M[i] \cdot u$ is equal to 0 (over $\bbr$),
and otherwise $x_i \defeq 0$.
Since all the column vectors whose indices are in the support of 
$x \flws \cald_0$
lie in a random linear space of dimension $n-1$, the function $f$ must reject
$\cald_0$.

To obtain a sunflower bound for $\cald_0$,
we would like to show that $\cald_0$ is 
$(t,\eps)$-pseudo-independent and apply \Cref{thm:sf-intro}.
Every column in our construction of $M$
will have a support of size $s = n^{\Omega(1)}$.
We use the design property to show that,
for any set $S = \set{c_1,\dots,c_t}\subseteq [m]$
of at most $t = s^{0.1}$ columns of $M$,
the set 
\[R = \bigcup_{i\neq j} (\supp(M[c_i]) \cap \supp(M[c_j]))\]
has size at most $s^{0.2}$.
Thus, for any $b \in \set{-1,+1}^R$, conditioning on the event
$\set{u_R = b}$ gives us that
that the random bits $x_{c_i},\dots,x_{c_t}$ 
are independent
(where $x \flws \cald_0$).
This means that $(\cald_0)_S$ is a convex combination of product distributions and
thus
$\cald_0$ is pseudo-independent.
It only remains to estimate the marginal probabilities,
but this can be done
using standard calculations about one-dimensional random walks
(sums of Rademacher random variables).

In our actual constructions, we will only be able to show Properties (1) and
(2) for a subset $C \sseq [m]$ of the columns such that $C =
\Omega(m)$. The arguments above can then be easily adapted to show a
lower bound for a restriction $g$ of $f$ that is defined only on the
columns of $C$.

\paragraph{Explicit construction of $M$.}
The main challenge in the explicit construction
is to balance the design property as well as the linear-algebraic expansion
properties. 
Informally, the columns of the
matrix are indexed by low-degree univariate polynomials over a finite field
$\mathbb{F}_q$, and the rows are essentially indexed by $[q] \times [s]$,
where $s$ is the sparsity parameter. The $((i,j), f)$-th entry is equal to
$1$ if and only if $f(j) = i$. Thus, the columns essentially encode the
graph of the univariate polynomials $f$. The limited intersection property
of the columns follows immediately from the low degree of the polynomials.
However, the main novelty lies in proving the linear-algebraic expansion
property. To this end, the key technical lemma we prove states that the
space spanned by matrices obtained from the graphs of all functions is
equal to the space spanned by matrices obtained from the graphs of
low-degree polynomials with affine shifts.
For the full technical details, we refer to Section \ref{sectExplicitConstructionRealCase}.

\subsection{Conclusions and Open Problems}
\label{sec:open}

\paragraph{Improved derandomized sunflowers.}

One of our main results is a robust sunflower bound for
pseudo-independent distributions, which
also achieves better seed-length in the setting of bounded independence.
Our results go beyond what is possible with the best known bounds for fooling DNFs using $k$-wise independent distributions~\cite{Tal17},
and it is natural to
ask whether they are tight:
\begin{question}
Is it possible to show better bounds on the size of 
width-$w$
monotone 
DNFs without
$(\cald,\delta)$-sunflowers, where $\cald$ is $t$-wise independent with
marginal probability $q$?
\end{question}
Inspired by the optimal robust sunflower
bounds~\cite{ALWZ21,Rao2020,BellChueluechaWarnke2021},
we conjecture that the right bound for width-$w$ DNFs is
$O(\frac{1}{q} \cdot \log(w/\delta))^w$,
for some $t = \widetilde{O}(w + \log(1/\delta))$.
If this is true, 
then we can obtain $2^{\tilde{\Omega}(\sqrt{n})}$ monotone Boolean circuit lower bounds
for a Boolean function in $\P$ (in fact, even in $\NC$),
matching the best known lower bounds for a function in
$\NP$~\cite{CKR22}.
Hopefully some of our ideas can a provide a starting point
for this direction.

\paragraph{New applications of derandomized sunflowers.}
As an application of our derandomization of sunflowers and our explicit construction via Nisan-Wigderson designs,
we significantly improved known separations between general arithmetic circuits and the radical of monotone circuits.
Our bounds are obtained via the framework of~\cite{CFMSY25}, which established a connection between sunflower bounds for weakly-independent distributions, matrices with pseudorandom properties, and monotone Boolean lower bounds.
Given the wide range of applications for sunflowers in theoretical computer science,  we ask whether our new bounds can be used to improve other applications:
\begin{question}
Can derandomized sunflower bounds find applications in other contexts where
robust sunflowers are typically applied?
In particular, could they
have impact on lifting theorems in communication complexity~\cite{DBLP:conf/innovations/LovettMMPZ22,Kelley21}?
\end{question}

\paragraph{Further radical lower bounds.}

This paper introduced the notion of the radical $\sqrt{\Ccal}$ of a complexity class $\Ccal$, which was briefly mentioned in the work of Hrube\v{s} and Yehudayoff~\cite{HY21}.
This concept is also interesting from the (nonmonotone) point of view of white-box PIT algorithms, factoring, and algebraic hardness vs. randomness.
For these settings, the most interesting regime is when the degree of the powers are bounded by a polynomially growing function on the number of variables,
and many known closure of factoring results imply that the bounded-degree radical of a class $\Ccal$ is equal to $\Ccal$~\cite{Kaltofen87, BKRRSS26}.
It is an open question is this area to understand whether
the same closure results are true when the degree can be bounded by an exponentially growing function in the number of variables.
For example, see Conjecture 8.3 in~\cite{Burgisser00} and the historical overview in~\cite{DuttaSaxenaS18}.
We believe it would be interesting to understand the relationship between 
$\sqrt{\Ccal}$ and $\Ccal$ for nonmonotone classes,
\bnote{I think even for monotone circuits this is open and interesting, but
the deadline is soon so I won't write anything.}
specially with respect to the algorithmic prospects of the notion of radical mentioned in the introduction of this paper.

For the monotone setting, we ask whether our separations can be improved:
\tnote{As Bruno asked in his email, do we know whether $\sqrt{\textbf{mVP}} = \textbf{mVP}$, that is, a closure of factoring for monotone classes? I guess we are interested in lower bounds}
\begin{question}
Can we improve our lower bounds for the radical of monotone arithmetic
circuits?
In particular, is there a technique for proving such bounds that
does not rely on monotone Boolean lower bounds?
\end{question}
The main motivation for the second question is to avoid the long-standing
problem of proving $2^{\Omega(n)}$ lower bounds for monotone Boolean
circuits computing a function in $\NP$, and the potential limitations of
the known techniques for proving monotone Boolean circuit lower
bounds~\cite{CKR22}.

The motivation for the first question is to understand whether there is a radical analogue of the $2^{\Omega(n)}$-lower bound for monotone circuits computing a polynomial computed by depth-3 circuits obtained by~\cite{CFMSY25}. %
This question is related to common hardness escalation phenomena in algebraic complexity:
a priori, it could be the case that any depth-3 arithmetic circuit of size $n^{O(1)}$ computing an $n$-variate monotone polynomial $P$ has a power $k\in\Naturals$ such that $P^k$ has a monotone circuit of size $2^{O(\sqrt{n})}$.
This would be a quite surprising result,
however many of the known lower bound techniques in algebraic complexity consist in lifting a lower bound for a restricted model of computation to a more general model via a nontrivial relationship between the complexity measures. 

Note that our superpolynomial separation between constant-depth arithmetic
circuits and the radical of monotone circuits rules out the existence of a
polynomially-sized conversion of constant depth circuits into the radical of
monotone circuits.

\subsection{Organization of the Paper}

The rest of the paper is organized as follows.
Section~\ref{sectRobustSunflower} proves our derandomization of robust sunflower lemma via the method of moments.
Section~\ref{sectImprovedSeparations} uses our improved sunflower bounds to obtain better separation between nonmonotone arithmetic circuits and the radical of monotone circuits.
Section~\ref{sectExplicitConstructionRealCase} analyses a linear-algebraic expansion property of Nisan-Wigderson designs, which is used in Section~\ref{sectImprovedSeparations} to obtain an explicit construction of our nonmonotone circuit upper bounds.

\section{Robust Sunflower Bounds via the Method of Moments}
\label{sectRobustSunflower}

The main goal of this section is to obtain sunflower bounds (Corollary~\ref{cor:robust-sf-main}) for pseudo-independent distributions. We start by recalling a few definitions.

\begin{definition}
    For $r > 1$, we say that a DNF $\Fcal$ with $N$ terms is \emph{$r$-spread}
    if, for any subset $S$ of variables,
    the number of terms of $\Fcal$ containing all the variables from $S$ is at most $N\cdot r^{-|S|}$.
    For a term $F$ of the DNF $\Fcal$, we use $\mathrm{Vars}(F)$ to denote the set of variables appearing in $F$.
    We say $\Fcal$ is \emph{monotone} if every term only has positive literals.
\end{definition}

Note that throughout this paper we interpret set systems over $[n]$ as monotone DNFs over the variables $x_1,\dotsc, x_n$, where each set $S$ in the system corresponds to a term $\Land_{i\in S} x_i$ in the DNF.

\begin{definition}
    We say that a distribution $\mathcal{D}$ on $\{0,1\}^n$ is
    \emph{$(t,\varepsilon)$-pseudo-independent with marginal probability $q$}
    if, for any set $S\subseteq [n]$ of at most $t$ coordinates,
    the marginal distribution $\mathcal{D}_S$ on $\{0,1\}^S$ is a convex combination of distributions,
    each of which is a product distribution on $\{0,1\}^S$
    with the marginal distribution on each coordinate
    $i\in S$ being a Boolean random variable that is set to $1$ with probability $q'_i \in [q\cdot (1-\varepsilon), q\cdot (1+\varepsilon)]$.
\end{definition}

We now state the main combinatorial theorem of this paper.

\begin{theorem}[Spread lemma for pseudo-independent distributions]
    \label{thm:robust-sunflower-pseudo}
    The following holds for every $0<q,\varepsilon<1$ and $w,t\in\Naturals$ and $r > 1$
    as long as $w\leq t$ and $\varepsilon \leq \gamma/w$ for small enough constant $\gamma > 0$.
    Let $\Fcal$ be an $r$-spread monotone $w$-DNF on $n$ variables with $N$ %
    terms $F_1, \dotsc, F_N$, each having width exactly $w$.
    If $\mathcal{D}$ is an $(t,\varepsilon)$-pseudo-independent
    distribution with marginal probability $q$,
    then
    \[
    \prob{x\sim \mathcal{D}}{\Fcal(x) = 0}\leq 
    \exp(-\Omega(\min\{t/w,rq/w\})),
    \]
    where $\Omega(\cdot)$ hides an absolute constant.
\end{theorem}

\begin{remark}
\label{rem:non-monotone}
The only place we use monotonicity is where we argue that each term is
satisfied on a random input with probability $q^w$. This also holds for
\emph{non-monotone} DNFs in the case $q = 1/2$. Thus our proof also works
in that setting. We omit the details.
Alternatively, a reduction can be found
in~\cite[Claim 2.3]{DBLP:conf/innovations/LovettMMPZ22}.
\end{remark}

We present a proof overview of Theorem~\ref{thm:robust-sunflower-pseudo}
in Section~\ref{sectProofIdeaSpreadnessLemmaPIndDist},
followed by its proof in Section~\ref{sectProofOfSpreadnessLemmaPIndDist}.
Then, in Section~\ref{sectSunflowerBoundsPIndDist}, we obtain sunflower bounds for pseudo-independent distributions as a corollary of Theorem~\ref{thm:robust-sunflower-pseudo}.

\subsection{Proof Idea} 
\label{sectProofIdeaSpreadnessLemmaPIndDist}

We want to upper bound
\[\prob{x}{\Fcal(x) = 0}\]
for $x$ being a element of $\bcube{n}$ sampled according to $\mathcal{D}$. 
We will prove this by computing a high enough moment of an associated random variable.

Assume that $x$ is sampled according to distribution $\mathcal{D}$. For every $i\in [N]$, let 
\[
    Y_i(x)\defeq 1_{\set{F_i(x) = 1}},\text{ and }
    \alpha\defeq q^w,\text{ and }
    X_i\defeq Y_i-\alpha,
\]
and let 
\begin{equation}
\label{eqDefRandomVarsAndMean}
  X\defeq \sum_{i\in [N]} X_i,\text{ and }
  Y\defeq \sum_{i\in [N]} Y_i,\text{ and }
  \mu\defeq N\alpha.
\end{equation}
For every even $p\in\Naturals$, we know that
\[\prob{x}{\Fcal(x) = 0}
= \prob{x}{X = -\mu}
\leq \prob{x}{X^p \geq \mu^p}
\leq \frac{\Exp[X^p]}{\mu^p},\]
so all we have to do is to obtain a good estimate for the $p$-th moment $\Exp[X^p]$ of $X$. 
We will obtain such an estimate in two steps.
\begin{enumerate}
    \item For any $s\in\Naturals$ and $i_1,\ldots, i_s\in [N]$, define
    \[
    S_{i_1,\ldots, i_s} := \left|\Exp\left[\prod_{j=1}^s X_{i_j}\right]\right|.
    \]
   We will first obtain, for any $i_1,\ldots, i_p\in [N]$, an upper bound
   on $S_{i_1,\ldots, i_p}$ based on the dependency structure of the
   underlying terms. This requires the underlying distribution
   $\mathcal{D}$ to be $pw$-wise independent
   (eventually, we will fix $p
   \leq t/w$, so $(t,\varepsilon)$-pseudo-independence is enough).
    \item Bound 
        \[\Exp[X^p] = \sum_{i_1,\dotsc,i_p\in [N]} \Exp\left[\prod_{j=1}^s X_{i_j}\right]\]
        by a suitable covering argument applied to all the terms of the above form $S_{i_1,\dotsc,i_p}$.
        In particular, we partition the terms $S_{i_1,\ldots, i_p}$ using the partition of $[p]$ induced by the dependency graph corresponding to the terms $F_{i_1},\ldots, F_{i_p}$.  This is where the $r$-spreadness of $\Fcal$ will be used, along with a  careful analysis of the number of such partitions with specified statistics.
\end{enumerate}

\subsection{Proof of Theorem~\ref{thm:robust-sunflower-pseudo}}
\label{sectProofOfSpreadnessLemmaPIndDist}

We now begin the formal argument with an execution of Step 1 in the above proof sketch.
Let 
\[p\defeq ct/w\]
for a sufficiently small constant $c\in\Reals$ to be defined later.

\begin{definition}
For any $(i_1,\ldots, i_s )\in [N]^s$,
let the \emph{dependency graph} $G_{i_1,\ldots,i_s}$ of $(i_1,\ldots, i_s)$
be an undirected graph with vertex set $[s]$ and
edges between distinct $a,b\in [s]$ if and only if  $F_{i_a}$ and $F_{i_b}$ share a variable.
\end{definition}

\begin{lemma}[A basic bound for $S_{i_1,\ldots,i_s}$]
    \label{lem:connGpseudo}
    Assume $p \leq t/w$ and $s\in [p].$ Fix any $(i_1,\ldots, i_s)\in [N]^s.$ %
    The following hold:
    \begin{itemize}
        \item If $s = 1,$ then $S_{i_1}\leq \alpha \cdot 2\varepsilon w.$ 
        \item If $1 < s \leq p$,
            then
            \[
                S_{i_1,\ldots, i_s} \leq \alpha^s \cdot 4^{s} \cdot \prod_{j=1}^s \left(\frac{1}{q}\right)^{|\mathrm{Vars}(F_{i_j})\cap \mathrm{Vars}(F_{i_1,\ldots,i_{j-1}})|},
            \]
            where $\mathrm{Vars}(F_{i_1,\ldots,i_{j-1}})$ denotes
            $ \mathrm{Vars}(F_{i_1})\cup \mathrm{Vars}(F_{i_2})\cup \cdots \cup \mathrm{Vars}(F_{i_{j-1}})$ if $j > 1$,
            and $\emptyset$ otherwise.
    \end{itemize}
\end{lemma}

\begin{proof}[Proof of Lemma~\ref{lem:connGpseudo}]
   We start with the case $s = 1$. In this case
   \begin{equation}
   \label{eq:Si1}
   	      S_{i_1} = |\Exp[X_{i_1}]| = \left| \prob{x\sim \mathcal{D}}{Y_{i_1} = 1} - \alpha\right|
   \end{equation}
    Since $t\geq w = |\Vars(F_{i_1})|$ and the distribution $\mathcal{D}$ is $(t,\varepsilon)$-pseudo-independent, it suffices to bound the above expression for each product distribution $\mathcal{D}'$ on $\{0,1\}^{\mathrm{Vars}(F_{i_1})}$ where each coordinate $a$ is set to $1$ with probability $q_a\in [q(1-\varepsilon), q(1+\varepsilon)]$.
    For such a $\mathcal{D}'$, it follows that
    \[
    \prob{x\sim \mathcal{D}'}{Y_{i_1} = 1}
    \leq q^w (1+\varepsilon)^w
    \leq \alpha\cdot (1+2\varepsilon w)
    \]
    since $\varepsilon  \leq \gamma/w$ and the constant $\gamma$ is small enough.
    Similarly, we also get 
    \[\prob{x\sim \mathcal{D}'}{Y_{i_1} = 1} \geq \alpha\cdot (1-2\varepsilon w).\]
    By plugging these bounds into (\ref{eq:Si1}), we obtain that $S_{i_1} \leq \alpha \cdot 2\varepsilon w$. 
    
    From now on, we assume that $s \geq 2$. For notational simplicity, we assume $i_j = j$ for each $j\in [s]$.
	We note that 
	\begin{align*}
	S_{1,\ldots, s}
     &=  \left|\Exp\left[\prod_{j=1}^s (Y_{j}-\alpha)\right]\right|
      \leq \sum_{T\subseteq [s]}\alpha^{s-|T|} \cdot \left|\Exp\left[ \prod_{j\in T} Y_{j}\right] \right|\\
     &=\sum_{T\subseteq [s]}\alpha^{s-|T|} \cdot\prob{x\sim \mathcal{D}}{\bigwedge_{j\in T}F_{j}(x) = 1}\\
     &\leq 2^s \cdot \max_{T\subseteq [s]} \left(\alpha^{s-|T|} \cdot\prob{x\sim \mathcal{D}}{\bigwedge_{j\in T}F_{j}(x) = 1}\right).
	\end{align*}
	Fix any $T \defeq \{j_1,\ldots, j_t\}\subseteq [s]$.
    Since $\mathcal{D}$ is $(t,\varepsilon)$-pseudo-independent and 
    \[t\geq pw \geq sw \geq \left|\bigcup_{j\in T}\mathrm{Vars}(F_j)\right|,\]
    it suffices to bound the above for a product distribution $\mathcal{D}'$ on $\{0,1\}^{\bigcup_{j\in T}\mathrm{Vars}(F_j)}$ where each coordinate $a$ is set to $1$ with probability $q_a\in [q(1-\varepsilon), q(1+\varepsilon)]$. For such a distribution $\mathcal{D}'$, we have
	\begin{align}
	\prob{x\sim \mathcal{D}'}{\bigwedge_{j\in T} F_{j}(x) = 1}  &\leq (q(1+\varepsilon))^{|\bigcup_{j\in T}\mathrm{Vars}(F_j)|}\label{eq:condprob}\\
    &\leq  (1+\varepsilon)^{tw}\cdot q^{|\bigcup_{j\in T}\mathrm{Vars}(F_j)|}\notag\\
	&\leq 2^t \cdot q^{|\bigcup_{j\in T}\mathrm{Vars}(F_j)|}
    \label{eqBoundWhenSmallGamma}\\
	 &= 2^t\cdot q^{\sum_{k=1}^t |\mathrm{Vars}(F_{j_k})| - |\mathrm{Vars}(F_{j_k})\cap \mathrm{Vars}(F_{j_1,\ldots,j_{k-1}})|}\notag\\
	& = 2^t q^{tw} \cdot \prod_{k=1}^t\left(\frac{1}{q}\right)^{|\mathrm{Vars}(F_{j_k})\cap \mathrm{Vars}(F_{j_1,\ldots,j_{k-1}})|}\notag\\
	&\leq  
    2^t \alpha^t \cdot
    \prod_{j=1}^s\left(\frac{1}{q}\right)^{|\mathrm{Vars}(F_{j})\cap
    \mathrm{Vars}(F_{1,\ldots,j-1})|}\notag\\
    &\leq     2^s \alpha^t \cdot
    \prod_{j=1}^s\left(\frac{1}{q}\right)^{|\mathrm{Vars}(F_{j})\cap
    \mathrm{Vars}(F_{1,\ldots,j-1})|}\notag,
\end{align}
    where (\ref{eqBoundWhenSmallGamma}) follows from
    $(1+\varepsilon)^w \leq (1+2\varepsilon w) \leq (1+2\gamma)\leq 2$ as above for $\gamma$ small enough.
Since we bounded the above for any product distribution $\mathcal{D}'$ with the prescribed marginals, we have the same bound for the distribution $\mathcal{D}$. Plugging this bound into the upper bound for $S_{i_1,\ldots, i_s}$ above yields the lemma.
\end{proof}

We now extend the bound from Lemma~\ref{lem:connGpseudo} to sum over all $(i_1,\ldots, i_s)\in [N]^s$ with a connected dependency graph. This is where we will use the spreadness of the DNF.

\begin{lemma}[The sum over connected $(i_1,\ldots,i_s)$]
    \label{lem:sumconnGpseudo}
    Assume $p \leq \min\{t/w,qr/10w\}$. 
    The following hold:
    \begin{itemize}
        \item If $s=1$, then
            \[
                \sum_{i_1\in [N]^s} S_{i_1}\leq N\alpha\cdot 2\cdot \varepsilon w.
            \]
        \item If $1 < s\leq p$, then
            \[
                \sum_{\substack{(i_1,\ldots,i_s)\in [N]^s\\ \text{$G_{i_1,\ldots,i_s}$ connected}}} 
                S_{i_1,\ldots,i_s} \leq (N\alpha)^s\cdot 2^{O(s)}\cdot \left(\frac{sw}{rq}\right)^{s-1}.
            \]
    \end{itemize}
\end{lemma}

\begin{proof}[Proof of Lemma~\ref{lem:sumconnGpseudo}]
The case $s = 1$ is immediate from Lemma~\ref{lem:connGpseudo} above since there are at most $N$ choices for $i_1$. For the rest of the lemma we assume that $s > 1$.
For a given $(i_1,\ldots, i_s),$ define the \emph{intersection pattern} $L_{i_1,\ldots, i_s}$ of $(i_1,\ldots, i_s)$ to be
\[L_{i_1,\ldots, i_s} \defeq  (\ell_2,\ldots, \ell_s),\]
where $\ell_j \defeq |\mathrm{Vars}(F_{i_j})\cap \mathrm{Vars}(F_{i_1,\ldots, i_{j-1}})|.$ 

We will partition the sum in the lemma statement based on the intersection patterns.
For any fixed intersection pattern $L \defeq (\ell_2,\ldots, \ell_s)$,
we have by Lemma~\ref{lem:connGpseudo} above that 
\begin{equation}
\label{eq:fixedintpattern}
\sum_{\substack{(i_1,\ldots,i_s)\\ L_{i_1,\ldots, i_s} =L}} S_{i_1,\ldots,i_s}
\leq \alpha^s\cdot 4^{s}\cdot \left(\frac{1}{q}\right)^{\ell_2+\cdots + \ell_s}\cdot N(L)
\end{equation}
where $N(L)$ denotes the number of $(i_1,\ldots, i_s)\in [N]^s$ such that $L_{i_1,\ldots, i_s} = L$. 

We can bound $N(L)$ using the spreadness of the DNF $\Fcal$. 
More precisely, given $L$, the number of ways of choosing an $(i_1,\ldots, i_s)$ such that $L_{i_1,\ldots, i_s} = L$ is at most
\[
N\cdot \prod_{j=2}^s\left( \binom{(j-1)w}{\ell_j}\cdot N\cdot r^{-\ell_j} \right)
\]
where:
\begin{itemize}
    \item the first $N$ is the number of ways of choosing $i_1$, and 
    \item for each $j> 1$, the $j$th term in the product is the number of ways of choosing $i_j$ given $i_1,\ldots, i_{j-1}$.
    Here, we use the spreadness of $\Fcal$:
    to choose $i_j$, we can first fix some subset $V$ of size $\ell_j$ of $\mathrm{Vars}(F_{i_1})\cup \cdots \cup \mathrm{Vars}(F_{i_{j-1}})$, which can be done in at most $\binom{(j-1)w}{\ell_j}$ ways;
    and then choosing a term that contains $V$,
    which can be done in at most $N\cdot r^{-\ell_j}$ ways using the spreadness property of $\Fcal$.
\end{itemize}
Plugging the above bound into
(\ref{eq:fixedintpattern}) we get
\begin{align*}
    \sum_{\substack{(i_1,\ldots,i_s)\in [N]^s:\\ L_{i_1,\ldots, i_s} =L}}
    S_{i_1,\ldots,i_s} &\leq N^s\alpha^s\cdot 4^{s}\cdot
    \left(\frac{1}{q}\right)^{\ell_2+\cdots + \ell_s}\cdot
    \prod_{j=2}^s\left( \binom{(j-1)w}{\ell_j}\cdot r^{-\ell_j} \right)\\
    &\leq N^s\alpha^s\cdot 4^{s}\cdot \prod_{j=2}^s\left(\frac{1}{q^{\ell_j}} \cdot ((j-1)w)^{\ell_j}\cdot r^{-\ell_j} \right)\\
    &=  N^s\alpha^s\cdot 4^{s}\cdot \prod_{j=2}^s\left(\frac{(j-1)w}{qr}\right)^{\ell_j}.  
\end{align*}
Moreover, for $v\defeq \sum_{i=2}^s \ell_i$, we have
\begin{align*}
    \sum_{\substack{(i_1,\ldots,i_s)\in [N]^s:\\
          L\leq L_{i_1,\ldots,i_s}\eqdef (v_2,\dotsc, v_s)}}  S_{i_1,\ldots,i_s}
    &\leq N^s\alpha^s\cdot 4^{s}\cdot
    \sum_{L\leq (v_2,\ldots, v_s)\in\Naturals^s} \prod_{j=2}^s\left(\frac{(j-1)w}{qr}\right)^{v_j}\\
    &\leq N^s\alpha^s\cdot 4^{s}\cdot
    \prod_{j=2}^s \sum_{v_j=\ell_j}^\infty \left( \frac{(j-1)w}{qr}\right)^{v_j}\\
    &\leq N^s\alpha^s\cdot 4^{s}\cdot \prod_{j=2}^s\left(\frac{2(j-1)w}{qr}\right)^{\ell_j} \\
    &\leq N^s\alpha^s\cdot 4^{s}\cdot \left(\frac{2sw}{qr}\right)^{\sum_{i=2}^s \ell_i}\\
    &\leq N^s\alpha^s\cdot 4^{s}\cdot \left(\frac{2sw}{qr}\right)^{v}.
\end{align*}
where the third inequality used the assumption that $s\leq p\leq qr/10w$ to deduce that each term of the product is a geometric progression with common ratio less than $1/2$. 

Note that, for every $i\in [N]^s$ such that $G_i$ is connected, 
the intersection pattern $L_i\in\Naturals^s$ of $i$ satisfies
\[L_i \geq L
    \text{ for some
    $L\in\setbar*{(\ell_2,\dotsc,\ell_s)\in\Naturals^s}{\sum_{i=2}^s\ell_i = s-1}\eqdef \Lcal_{\text{min}}$,}
\]
since every connected graph contains a spanning tree as subgraph and
the patterns in $\Lcal_{\text{min}}$ correspond to patterns of spanning trees.
Furthermore,
\[\card{\Lcal_{\text{min}}} = \binom{s-1+s-1-1}{s-1}\leq 2^{O(s)}.\]
Hence, we obtain
\begin{align*}
    \sum_{\substack{(i_1,\ldots,i_s)\in [N]^s:\\
          \text{$G_{i_1,\dotsc,i_s}$ connected}}} S_{i_1,\ldots,i_s}
    &\leq \sum_{(\ell_2,\ldots,\ell_s)\defeq L\in \Lcal_{\text{min}}}
        \sum_{\substack{(i_1,\ldots,i_s)\in [N]^s:\\
        L\leq L_{i_1,\dotsc,i_s}}} S_{i_1,\ldots,i_s} \\
    &\leq 2^{O(s)}\cdot
        \left(N^s\alpha^s\cdot 4^{s}\cdot\left(\frac{2sw}{qr}\right)^{s-1}\right) \\
    &\leq N^s\alpha^s\cdot 2^{O(s)}\cdot\left(\frac{sw}{qr}\right)^{s-1}
\end{align*}

\bnote{Check if we actually need the extra $s!$.}
\snote{If we can avoid the extra $s!$, we would get $\exp(-\Omega(r/w))$.}
\tnote{PLEASE CHECK WHETHER THE PROOF ABOVE IS RIGHT.}
\end{proof}

\begin{definition}
    Let $(i_1,\ldots, i_p)\in [N]^p$ be a tuple with a possibly disconnected dependency graph
    $G_{(i_1,\ldots,i_p)}$.
    For an unordered partition of $\pi \defeq \{A_1,\ldots, A_m\}$ of the set $[p]$,
    we say that the tuple $(i_1,\ldots, i_p)$ is \emph{$\pi$-consistent}
    if the connected components of $G_{(i_1,\ldots,i_p)}$ are $A_1,\ldots, A_m$.
    We use $\mathcal{C}_\pi$ to denote the set of $\pi$-consistent tuple $(i_1,\ldots, i_p)$.
    For a set $A\defeq \set{j_1,\dotsc, j_s}\subseteq [p]$ with $j_1 < \cdots < j_s$,
    we use $i_A$ to denote the tuple $(i_{j_1},\ldots, i_{j_s})$.
\end{definition}

\begin{lemma}[Sum over $\pi$-consistent $(i_1,\ldots, i_p)$]
\label{lem:piconsistentpseudo}
    Assume $p \leq  \min\{t/w,qr/10w\}$.
    Fix any unordered partition $\pi \defeq \{A_1,\ldots, A_m\}$ of $[p]$,
    and let $a_j\defeq |A_j|$ for each $j\in [m]$.
    Let $c_1$ denote the number of singletons in $\pi$, that is,
    number of $j\in [m]$ such that $a_j = 1$.
    Then, we have
	\[
	\sum_{(i_1,\ldots, i_p)\in \mathcal{C}_\pi} S_{i_1,\ldots, i_p}
    \leq \mu^p \cdot 2^{O(p)} \cdot (\varepsilon w)^{c_1}\cdot  \left(\frac{pw}{rq}\right)^{p-m}
    \]
	where $\mu = N\alpha$ (as in Equation~\ref{eqDefRandomVarsAndMean}).
\end{lemma}

\begin{proof}[Proof of Lemma~\ref{lem:piconsistentpseudo}]
Without loss of generality, we may assume that $a_j =1$ for $j\leq c_1$ and $a_j > 1$ for $j > c_1$.

Fix any $(i_1,\ldots, i_p)\in \mathcal{C}_\pi$.
Note that
\[S_{i_1,\ldots, i_p} = S_{i_{A_1}}\cdots S_{i_{A_m}}\]
by the independence of the variables $(X_{i_j}: j \in A_{k})$ and $(X_{i_{j'}}: j'\in A_{k'})$ for any $k\neq k'$. 
Using independence across $A_1,\ldots, A_m$ and using Lemma~\ref{lem:sumconnGpseudo},
we see that %
\begin{align}
\sum_{(i_1,\ldots, i_p)\in \mathcal{C}_\pi} S_{i_1,\ldots, i_p}
&=\sum_{\substack{(i_1,\dotsc, i_p)\in [N]^p\\ \text{$G(i_{A_j})$ connected $\forall j$}}} S_{i_{A_1}}\cdots S_{i_{A_m}}\notag\\
&= \prod_{j=1}^{c_1}
\left( \sum_{i_j\in [N]} S_{i_j}\right) \cdot
\prod_{j=c_1+1}^m \left(\sum_{\substack{i_{A_j} \in [N]^{A_j}\\ \text{$G(i_{A_j})$ connected}}} S_{i_{A_1}}\right)\notag\\
&\leq (N\alpha)^{\sum_{j\in [m]}a_j}\cdot
      2^{O(\sum_{j\in [m]}a_j)}
      \cdot (\varepsilon w)^{c_1}
      \left(\frac{a_jw}{rq}\right)^{\sum_{j > c_1} (a_j-1)}\notag\\
    &=(N\alpha)^p\cdot 2^{O(p)}\cdot (\varepsilon w)^{c_1} \cdot  \left(\frac{pw}{rq}\right)^{p-m}\notag.
\end{align}
Using the fact that $N\alpha = \mu$, we obtain the statement of the lemma.
\end{proof}

Now we have all the auxiliary lemmas to prove Theorem~\ref{thm:robust-sunflower-pseudo}.

\begin{proof}[Proof of Theorem~\ref{thm:robust-sunflower-pseudo}]
   For any even integer $p$,
\begin{equation}
\label{eq:moment-bound}
\prob{x}{\Fcal(x) = 0}
= \prob{x}{X = -\mu}
\leq \prob{x}{X^p \geq \mu^p}
\leq \frac{\Exp[X^p]}{\mu^p}.
\end{equation}
We will set $p$ to be the largest even number such that 
\begin{equation}
    \label{eqChoiceOfMoment}
    p \leq \min\{t/w, \eta(qr/w)\}
\end{equation}
for a small enough constant $\eta > 0$.

To bound $\Exp[X^p]$, we proceed as follows:
\begin{align*}
\Exp[X^p] &= \Exp\left[\sum_{(i_1,\ldots,i_p)\in [N]^p} X_{i_1}\cdots X_{i_p}\right]\leq \sum_{(i_1,\ldots,i_p)\in [N]^p} S_{i_1,\ldots,i_p}\\
&=\sum_{m=1}^p \sum_{c_1=\max\{0,m-p/2\}}^m \sum_{\substack{\pi = \{A_1,\ldots,A_m\}:\\ \text{$\pi$ has $c_1$ singletons}}} \sum_{(i_1,\ldots,i_p)\in \mathcal{C}_\pi}S_{i_1,\ldots,i_p},
\end{align*}
where we are using above that for any partition with $m$ parts, there must be at least $m - p/2$ parts of size $1$.

Applying Lemma~\ref{lem:piconsistentpseudo} to bound each inner sum, we obtain
\begin{align}
    \Exp[X^p]
    &\leq \sum_{m=1}^{p}\sum_{c_1=\max\{0,m-p/2\}}^m
          \sum_{\substack{\pi = \{A_1,\ldots,A_m\}:\\ \text{$\pi$ has $c_1$ singletons}}}
          \mu^p \cdot 2^{O(p)} \cdot \left(\frac{pw}{rq}\right)^{p-m}\cdot (\varepsilon w)^{c_1}\notag\\
    &\leq \mu^p 2^{O(p)}\cdot \sum_{m=1}^p \left(\frac{pw}{rq}\right)^{p-m} \cdot 
          \sum_{c_1=\max\{0,m-p/2\}}^m \left(\varepsilon w\right)^{c_1}\label{eq:Xpbound1}
\end{align}

We bound the above quantities based on the value of $m.$ First recall that $\varepsilon \leq \gamma/w$ for a small enough constant $\gamma > 0$. 

For $m\leq p/2$, we note that the inner sum is bounded by an absolute constant, so we obtain
\[
    \sum_{m=1}^{p/2} \left(\frac{pw}{rq}\right)^{p-m} \cdot 
    \sum_{c_1=\max\{0,m-p/2\}}^m \left(\varepsilon w\right)^{c_1}
    \leq O\left(\sum_{m=1}^{p/2} \left(\frac{pw}{rq}\right)^{p-m}\right)
    \leq O\left(\eta^{p/2}\right),
\]
where the last inequality follows from Equation~\ref{eqChoiceOfMoment}.
For $m > p/2,$ the inner sum is bounded by $O\left(\left(\varepsilon w\right)^{m-p/2}\right)$,
so we obtain
\begin{align*}
    \sum_{m=p/2+1}^{p} \left(\frac{pw}{rq}\right)^{p-m} \cdot 
    \sum_{c_1=\max\{0,m-p/2\}}^m \left(\varepsilon w\right)^{c_1}
    &\leq O\left(\sum_{m=p/2+1}^{p} \left(\frac{pw}{rq}\right)^{p-m}\cdot
                                    \left(\varepsilon w\right)^{m-p/2}\right)\\
    &\leq O\left(p\cdot (\eta+\gamma)^{p/2}\right)
\end{align*}
where we used the fact that $p\leq \eta(rq/w)$ and $\varepsilon \leq \gamma/w$ for the final inequality.

Using the above bounds in (\ref{eq:Xpbound1}), we obtain
\[
\Exp[X^p] \leq \mu^p 2^{O(p)}\cdot (\eta+\gamma)^{p/2} = \mu^p\cdot \exp(-\Omega(p))
\]
as long as $\eta,\gamma$ are small enough constants.
Plugging the above bounds
into (\ref{eq:moment-bound}), we obtain
\[
\prob{x}{\Fcal(x) = 0} \leq \exp(-\Omega(p))  = \exp(-\Omega(\min\{t/w,rq/w\})).
\]
\end{proof}

\subsection{Sunflower Bounds for Pseudo-Independent Distributions}
\label{sectSunflowerBoundsPIndDist}

Following~\cite{CFMSY25}, we have the following definition:

\begin{definition}
    \label{defRobustSunflower}
    We define a DNF $\mathcal{F}$ to be a \emph{$(\mathcal{D},\delta)$-sunflower} if 
    \[
        \prob{x\sim \mathcal{D}}{\mathcal{F}_\Delta(x) = 1} > 1-\delta
    \]
    where $\mathcal{F}_\Delta$ denotes the DNF obtained from $\Fcal$ by setting all literals that appear in \emph{all the terms} of $\Fcal$ to $1.$
\end{definition}

Theorem~\ref{thm:robust-sunflower-pseudo} along with a standard argument now implies the following sunflower bound.

\begin{corollary}[Theorem~\ref{thm:sf-intro}]
\label{cor:robust-sf-main}
There is an absolute constant $\theta_0$ so that the following holds.
Let
$r,q,w,\delta$ be such that $\delta \geq \exp\left(-\theta_0(rq/w)\right)$. 
Then,
there exist parameters
$t = \Theta(w\log(1/\delta))$ and
$\varepsilon = \Theta(1/w)$
such that the following holds.
For any
$(t,\varepsilon)$-pseudo-independent distribution $\mathcal{D}$ with marginals probability $q$,
any monotone DNF with at least $r^w$ terms of width exactly $w$ contains a
$(\mathcal{D},\delta)$-sunflower.
\end{corollary}
\begin{proof}
    Without loss of generality, assume the DNF $\calf$
    has exactly $N := r^w$ terms $F_1, \dots, F_N$ of
    width $w$.
    Let $\theta_0$ be the absolute constant in the 
    exponent of the
    concluding statement of \Cref{thm:robust-sunflower-pseudo}.
    For $t \defeq \frac{w\log(1/\delta)}{\theta_0}$,
    we obtain
    \begin{equation}
        \label{eq:constants}
        \min\set{t/w, rq/w} \geq \frac{\log(1/\delta)}{\theta_0},
    \end{equation}
    as $\delta \geq \exp(-\theta_0(rq/w)))$.

    If $\calf$ is $r$-spread, the result 
    follows immediately 
    from
    Inequality~(\ref{eq:constants})
    and \Cref{thm:robust-sunflower-pseudo}.
    Suppose then $\calf$ is not $r$-spread.
    Let $S$ be a maximal set witnessing a ``spreadness violation'',
    i.e., a set such that
    $\calf_S := \set{F_i \sm S : S \sseq F_i}$
    satisfies
    $\card{\calf_S} \geq r^{w-\card{S}}$.
    Note that $\calf_S$ is $r$-spread, otherwise
    $S$ would not be maximal.
    By Inequality~(\ref{eq:constants}) and \Cref{thm:robust-sunflower-pseudo},
    we get that $\calf_S$ is a $(\cald, \delta)$-sunflower,
    which implies that 
    $\set{F_i : S \sseq F_i}$ is a $(\cald,\delta)$-sunflower in $\calf$.
\end{proof}

\begin{remark}
\label{rem:non-monotone3}
     A near-identical argument also shows the same is true for non-monotone DNFs in the case $q=1/2.$
     Again check~\cite[Claim 2.3]{DBLP:conf/innovations/LovettMMPZ22} for details.
\end{remark}

\section{Improved Separations for the Radical of Monotone Circuits}
\label{sectImprovedSeparations}

The main theorem of this section is the following.
\begin{theorem}[Theorem~\ref{thm:radical-sep-intro}]
    \label{thm:radical-sep-main}
    There is an infinite sequence $(P_m(x_1,\ldots,x_m))_{m\in\Naturals}$ of  polynomials
    such that the following hold:
    \begin{enumerate}
        \item (Explicit Upper bound) $P_m$ is a monotone multilinear polynomial with an arithmetic circuit $C_m$ of size $m^{O(1)}$. Further, given $m$, the circuit $C_m$ can be computed in deterministic time $m^{O(1)}$.
        \item (Radical Lower bound) Any monotone arithmetic circuit computing a power of $P_m$ must have size at least $\exp(m^{\theExponent}).$
    \end{enumerate}
\end{theorem}

This improves the result of~\cite{CFMSY25} in two ways: it makes the upper bound uniform, and improves the separation from quasipolynomial to exponential. Furthermore, it allows us to obtain the first superpolynomial separation between nonmonotone constant-depth circuit and the radical of monotone circuits, via a subexponential depth-reduction for our upper bound.

\begin{corollary}[Corollary~\ref{cor:radical-ac0-intro}]
    For every constant $k \geq 1$,
    there is an infinite sequence $(Q_M(x_1,\ldots,x_M))_{M\in\Naturals}$
    of polynomials
    such that the following hold:
    \begin{enumerate}
        \item $Q_M$ is a monotone multilinear polynomial 
        with a 
        depth $O(k)$
        arithmetic circuit $C'_M$ of size $M^{O(1)}$. Further, given $M$, the circuit $C'_M$ can be computed in deterministic time $M^{O(1)}$.
        \item Any monotone arithmetic circuit computing a power of $Q_M$ must have size at least
        $M^{\Omega((\log M)^k)}$.
    \end{enumerate}
\end{corollary}
\begin{proof}
By known depth reduction techniques~\cite{AgrawalVinay08, Koiran12, Tavenas15, GKKS16}, for any constant $\Delta\geq 3,$
we can convert a circuit of size $m^{O(1)}$
computing a polynomial $P$ of degree $d$
into a depth-$\Delta$ circuit computing $P$ with size
\[2^{O(d^{1/(\Delta-1)}\cdot \log m)}.\]

Hence, the polynomial $P_m$ from Theorem~\ref{thm:radical-sep-main} can be computed by a depth $\Delta$ circuit of size
\[s'\leq 2^{O(m^{2/\Delta})} = M_\Delta^{O(1)},\]
for $M_\Delta\defeq 2^{m^{2/\Delta}}$.
Moreover, by the radical lower bound in Theorem~\ref{thm:radical-sep-main}, there is a constant $c>0$ such that any power of $P_m$ can only be computed by monotone circuits of size
\[2^{\Omega(m^c)}.\]
Let $\Delta\defeq \ceiling{4k/c}$ and $M\defeq M_\Delta$.
We define $Q_M(x_1,\dotsc, x_M)$ the padded version 
of $P_n$, that is,
\[Q_M(x_1,\dotsc, x_M)\defeq P_n(x_1,\dotsc, x_n).\]
Then $Q_M$ can be computed by a constant depth circuit of size $M^{O(1)}$,
 and any power of $Q_M$ can only be computed by monotone circuits of size
\[2^{\Omega(m^c)} = M^{\Omega(\log M)^k}\]
implying the desired lower bound.
\end{proof}

\paragraph{Organization.}
Our proof of \Cref{thm:radical-sep-main} is organized as follows.
In Section~\ref{sec:cfmsy}, we recall the basic framework of~\cite{CFMSY25}. We will use the same kind of `matrix-rank' polynomial sequences based on a sequences of matrices $(M_m)$ as in~\cite{CFMSY25} and thus our upper bounds will follow similarly. The  separation is proved by a monotone \emph{Boolean} circuit lower for an associated sequence of monotone Boolean functions $(f_m)$ and the proof of this is based on an abstract framework of~\cite{CFMSY25} for proving such lower bounds.

In Section~\ref{sec:propsM}, we identify sufficient properties of the matrix sequence $M_m$ that allow us to use the \cite{CFMSY25} lower bound framework to prove an exponential monotone Boolean circuit lower bound. We improve upon the quasipolynomial lower bound of~\cite{CFMSY25} because we are able to use our strong robust sunflower bounds (Corollary~\ref{cor:robust-sf-main}) for  pseudo-independent distributions while the lower bound of~\cite{CFMSY25} was based on the classical sunflower lemma for set-systems~\cite{ER60} which has worse parameters.

In Section~\ref{sec:constructM}, we state our results about constructing a sequence of matrices $(M_m)$ with the required properties from Section~\ref{sec:propsM} in deterministic time $\poly(m)$. We use this construction to finish the proof of Theorem~\ref{thm:radical-sep-main}. The details of the construction are given in Section~\ref{sectExplicitConstructionRealCase}.
Stated combinatorially, the columns of the matrices are a uniform family
of sets that satisfy a combinatorial design property and
furthermore satisfy a simple ``linear-algebraic expansion''
condition. We start with a well-known (see e.g.~\cite{Nisan91PRGs,NisanWigderson1994})  
construction of such set-systems based on graphs of low-degree
polynomials and show that they have the required linear-algebraic
properties. This latter fact may be independently interesting.

\subsection{The Framework of~\cite{CFMSY25}}
\label{sec:cfmsy}

\paragraph{Matrix-rank polynomials.} Given an infinite sequence of matrices $(M_m)$ where $M_m\in \mathbb{R}^{n\times m}$ with $n\leq m$ (here $n$ is a parameter that grows with $m$), we define a sequence of multilinear polynomials with non-negative real coefficients as follows. For each $m,$
\[
P_m(x_1,\ldots, x_m) \defeq \sum_{S\in \binom{[m]}{n}} \det(M_m[S])^2\cdot \prod_{i\in S}x_i
\]
where $M_m[S]$ denotes the $n\times n$ submatrix obtained by keeping only the columns indexed by $S$ (in increasing order). 

The following lemma was observed in~\cite{CFMSY25} using known constructions of arithmetic circuits computing the determinant~\cite{Berkowitz84,Mulmuley87}. While it is not explicitly stated, it is easy to see that the construction yields a uniform family of circuits as long as the matrix sequence is explicit.

\begin{lemma}[Upper bound from~\cite{CFMSY25}]
    \label{lem:alg-ubd}
    For any matrix $M_m$, the polynomial $P_m(x_1,\ldots,x_m)$ can be computed by an arithmetic circuit $C_m$ of size $\poly(m)$ and depth $O((\log m)^2).$ Furthermore, there is a deterministic algorithm that, given as input the matrix $M_m$, produces the circuit $C_m$ in time $\poly(m).$
\end{lemma}

\paragraph{Radical lower bounds via Boolean circuit lower bounds.} For a monotone polynomial $P\in \mathbb{R}[x_1,\ldots, x_m],$ we define a corresponding Boolean function $f_P:\{0,1\}^m\rightarrow \{0,1\}$ by 
\[
f_P(x) = 1 \Leftrightarrow P(x) > 0.
\]
For the monotone polynomial $P_m$ defined above, the corresponding Boolean function $f_m := f_{P_m}$ can be equivalently defined as
\[
f_m(1_S) = 1\Leftrightarrow \mathrm{rank}(M[S]) = n
\]
where $1_S\in\bcube{m}$ denotes the indicator vector of $S\subseteq [m].$

It is shown in~\cite{CFMSY25} that there is a sequence of matrices $(M_m)$ such that the monotone arithmetic complexity of any power of $P_m$ is superpolynomial. To do this, they use the following observation (see also Remark 6.5 from~\cite{HY21}).

\begin{remark}[\cite{CFMSY25}]
\label{obs:radicaltomonotone}
    If $P^k$ has a monotone arithmetic circuit of size $s$, then $f_P$ has a monotone Boolean circuit of size $s.$
\end{remark}

In particular, this observation reduces the problem of proving lower bounds on the monotone arithmetic circuit complexity of powers of $P_m^k$ to proving a lower bound on the monotone Boolean circuit complexity of the corresponding Boolean function $f_m.$ 

We will follow this strategy,
and we prove our monotone Boolean circuit lower bound
via an abstract framework for proving such lower bounds also from~\cite{CFMSY25}.
The framework is based on Razborov's approximation method~\cite{Raz85} and
followup works based on robust sunflowers~\cite{Rossman2014,CKR22,CGRSS25}.

\begin{theorem}[Framework for monotone Boolean circuit lower bounds~\cite{CFMSY25}]
    \label{thm:sunflower-gen-cfmsy}
    Let $m\in\Naturals$ and let $f : \{0,1\}^m \to \{0,1\}$ be a monotone function. Fix parameters $t,Q,r\in \mathbb{N}$.
    
    Assume that there exist probability distributions $\cald_0, \cald_1$ over $\{0,1\}^m$ such that
    \begin{enumerate}
        \item For any $T\subseteq [m]$ of size at most $t,$ 
        \[
        \prob{x\sim \mathcal{D}_1}{\bigwedge_{i\in T}x_i = 1} \leq Q^{-|T|},
        \]
        \item For some $w\leq t/2$ and any $\ell\in [2w]$,
        any monotone DNF with more than $r^\ell$ terms of width exactly $\ell$ contains a
        $(\mathcal{D}_0, \alpha m^{-3w})$-sunflower (Definition~\ref{defRobustSunflower}), where
        \[
    \alpha\defeq \min \set*{\Pr_{x \flws \cald_0}[f(x)=0],\; \Pr_{x \flws \cald_1}[f(x)=1]}. 
        \]
    \end{enumerate}
    
    If $8r\leq Q \leq r m$,
    then any Boolean monotone circuit computing $f$ has size at least
    \begin{equation*}
        \left( 
            \frac{c \alpha Q}{r} 
        \right)^w
    \end{equation*}
    where $c$ is a universal positive constant. %
\end{theorem}

\subsection{Monotone Lower Bounds from Properties of $(M_m)$}
\label{sec:propsM}

Fix a growing parameter $m$.
For brevity we will use $M,P,f$ to denote
the matrix $M_m$, the corresponding polynomial $P_m$ and the Boolean function $f_m$ respectively (as defined in Section~\ref{sec:cfmsy}).

In this section, we will show that if $M\in \mathbb{R}^{n\times m}$ has some suitable combinatorial and algebraic properties, then the corresponding Boolean function $f$ has high monotone Boolean circuit complexity.
These properties are related to the definition of well-behaved matrix used by~\cite{CFMSY25} to obtain their superpolynomial separation.

\paragraph{Well-behaved matrices.}
The following definition generalizes the notion of well-behaved matrix introduced by~\cite{CFMSY25}.
\begin{definition}
\label{defWellBehavedMatrixNewPaper}
Fix a subset $C\subseteq [m]$ and define $V\subseteq \mathbb{R}^n$ to be the subspace spanned by the columns of $M$ indexed by $C$. For integer parameters $s,k$, we say that a rank-$n$ matrix $M\in \mathbb{R}^{n\times m}$ is \emph{$(C,s,k)$-well-behaved} if the following conditions hold.
\begin{enumerate}
    \item (Design property)
        Each column $v\in\bcube{n}$ indexed by $C$ is a $\set{0,1}$-vector with
        support $\supp(v)\defeq\setbar{i\in [n]}{v_i\neq 0}$ of cardinality $s$.
    Furthermore, for any distinct $i,j\in C$, there are at most $k$ indices where the corresponding columns $M[i]$ and $M[j]$ are both $1.$
    \item (Linear-algebraic expansion inside $V$)
    The subspace $V$ is non-zero and, given any vector subspace $U$ of $V$, we have
    \[
    \prob{i\sim C}{M[i]\in U} \leq \frac{\dim(U)}{\dim(V)}.
    \]
\end{enumerate}
\end{definition}

The main theorem of this section is the following.

\begin{lemma}[Lower bounds from well-behaved matrices]
\label{lem:well-behaved}
Assume that $M$ is a $(C,s,k)$-well-behaved $n\times m$ matrix of rank $n$
with $s$ even, 
and that
\snote{Added evenness condition on $s$}
\bnote{It seems a lower bound on $|C|$ follows from the upper bound for
$s$, so the first condition is redundant.}
\[
\frac{s}{k^5} \geq m^{\Omega(1)},\text{ and }
s \leq \frac{|C|}{n\cdot m^{\Omega(1)}}.\]
Then, any monotone Boolean circuit computing $f$ has size at least $\exp(m^{\Omega(1)}).$
\end{lemma}

We will prove this theorem by applying the~\cite{CFMSY25} framework (Theorem~\ref{thm:sunflower-gen-cfmsy}) to a restriction $g$ of $f$ and suitable distributions $\mathcal{D}_0$ and $\mathcal{D}_1$. The proof follows the ideas set forth in \cite{CFMSY25}. The technical part of the section will be to show that the distribution $\mathcal{D}_0$ is an $(t,\varepsilon)$- pseudo-independent distribution for suitable parameters $r$ and $\varepsilon$ so that we can apply Corollary~\ref{cor:robust-sf-main}
(which is necessary to obtain the sunflower bound in the hypothesis of Theorem~\ref{thm:sunflower-gen-cfmsy}).

\paragraph{Restriction of $f$.}
We start by defining the restriction $g$ of $f$ that we will consider.
Fix $C,s,k$ as in the hypothesis of Lemma~\ref{lem:well-behaved}.
Note that $|C| \geq nm^{\Omega(1)}$.
As $M$ is rank $n$, there is a subset $C'$ of the columns of $M$ such that
$C'\cap C = \emptyset$ and the columns $C'$ and $C$ together span the entire space $\mathbb{R}^n.$ We fix the coordinates of the sampled point indexed by $[m]\setminus C$ as follows: $x_i = 1$ if $i\in C'$ and $x_i = 0$ if $i\in [m]\setminus (C\cup C').$ Under this setting, the function $f$ restricts to a monotone function $g$ on $m_1 := |C| = m^{\Omega(1)}$ variables. W.l.o.g. assume that $C = [m_1]$.

Clearly, the monotone Boolean circuit complexity of $g$ is at most that of $f$ and hence, it suffices to prove a lower bound for $g.$ 

Note also the following basic property of $g$, that follows directly from
its definition and the fact that the columns of $M$ indexed by $C$ span the
subspace $V$ of $\mathbb{R}^n.$
\bnote{Might be worth pointing out that, crucially, $C$ is not a basis
spanning $V$ -- it must contain several ``useless'' vectors, otherwise
$g$ is just the AND function.}
\begin{remark}
    \label{obs:g-defn}
    For any subset $S\subseteq [m_1]$,
    \[
    g(1_S) = 1 \Leftrightarrow \rank(M[S\cup C']) = n \Leftrightarrow \text{the columns of $M$ indexed by $S$ span $V$.}
    \]
\end{remark}

\paragraph{Hard distributions.}
We now define the distributions $\mathcal{D}_0$ and $\mathcal{D}_1$, which are defined similarly to the corresponding distributions in~\cite{CFMSY25}.

\begin{definition}
\label{defHardDistributions}
For the distribution $x\sim\mathcal{D}_1$:
let $\lambda \defeq 100 n \log n$,
\bnote{Since the size of $\card{C}$ has not been specified, how do you know
there are $n \log n$ columns inside of it?}
\tnote{We need $m_1=|C|$ to be larger than $\lambda$ so that Lemma~\ref{lem:spreadness-D1} makes sense}
and sample a set $I$ of $\lambda$ indices from $[m_1]$ independently and uniformly at random (i.e. $\lambda$ samples from $[m_1]$ with replacement);
and define $x\in\bcube{m_1}$ as $x_i\defeq 1$ for every $i\in I$, and $x_i\defeq 0$ for every $i\in [m_1]\drop I$.

For the distribution $x\sim\mathcal{D}_0$:
we sample a uniformly random $u\in \{-1,1\}^n$,
and define $x\defeq x(u)\in\bcube{m_1}$ as, for each $i\in [m_1]$, we set $x(u)_i = 1$ if and only if the inner-product $M[i]\cdot u$ is zero over the reals.
\end{definition}

We now prove that these distributions have the required properties to apply Theorem~\ref{thm:sunflower-gen-cfmsy}.

\subsubsection{Properties of $\mathcal{D}_1$}

Our proofs of Lemmas~\ref{lem:alpha-D1} and~\ref{lem:spreadness-D1} use essentially the same ideas from the corresponding lemmas in~\cite{CFMSY25}.
We postponed them to Appendix~\ref{appendixPropertiesOfD1}.

\begin{lemma}
\label{lem:alpha-D1}
    $
    \prob{x\sim \mathcal{D}_1}{g(x) = 1}\geq 1/2.
    $
\end{lemma}

\begin{lemma}
\label{lem:spreadness-D1}
    For any subset $T\subseteq [m_1]$, 
    $
    \prob{x\sim \mathcal{D}_1}{\bigwedge_{i\in T}x_i = 1} \leq \left(\frac{m_1}{\lambda}\right)^{-|T|}.
    $
\end{lemma}

\subsubsection{Properties of $\mathcal{D}_0$}

\begin{lemma}
\label{lem:alpha-D0}
    $
    \prob{x\sim \mathcal{D}_0}{g(x) = 0}\geq 1/2.
    $
\end{lemma}
\begin{proof}
First recall that, for every $i\in [m_1]$, we have that $M[i]$ is a Boolean vector of Hamming weight $s$,
and, for every $S\subseteq [m_1]$,
\[g(1_S) = 1 \iff \linspan\setbar{M[i]}{i\in [m_1]} = V\subseteq \linspan\setbar{M[i]}{i\in S}. \]
Thus,
\begin{align*}
    \prob{x\sim\Dcal_0}{g(x) = 1}
    &= \prob{x\sim\Dcal_0}{\linspan\setbar{M[i]}{i\in [m_1]}\subseteq \linspan\setbar{M[i]}{i\in [m_1], M[i]\cdot u = 0}} \\
    &\leq \prob{x\sim\Dcal_0}{\forall i\in [m_1], M[i]\cdot u = 0}
     \leq \prob{x\sim\Dcal_0}{M[1]\cdot u = 0},
\end{align*}
and this lemma follows from the fact that, for every $i\in [m_1]$,
\begin{align*}
\prob{x\sim\Dcal_0}{M[i]\cdot u = 0}
&= \frac{|\setbar{u\in\set{-1,1}^n}{M[i]\cdot u = 0}|}{2^n}  \\
&= 2^{-s}\card*{\setbar*{u\in\set{-1,1}^s}{\sum_{i=1}^s u_i = 0}} \\
&\leq \Theta(1/\sqrt{s}) \leq 1/2.
\end{align*}
\end{proof}

\begin{lemma}
    \label{lem:pseudo-ind-D0}
    For any $t\in\Naturals$ such that $(t^2k)^3 \leq s$,
    the distribution $\mathcal{D}_0$ is $(t,\Theta(1/s^{1/3}))$-pseudo-independent with marginal probability
    $q \defeq \Theta(1/\sqrt{s}).$
\end{lemma}
\begin{proof}
Let 
\[q\defeq \sqrt{\frac{2}{\pi s}}\text{ and } \eps\defeq cs^{-1/3}\]
for a constant $c$ to be defined at the end of this proof.
Let $S\subseteq [m]$ with $|S| = t$.
Our goal is to show that
the marginal distribution 
\[\Dcal\defeq (\Dcal_0)_S\]
of $\Dcal_0$ is a convex combination of distributions,
where each one of them is a product distribution on $x\in\bcube{S}$ with each coordinate $x_i$
being a Bernoulli random variable with success probability in the interval $[q(1-\eps), q(1+\eps)]$.

Let 
\[R\defeq \cup_{i,j\in S, i\neq j} \supp(M_i)\cap\supp(M_j),
\]%
where, by the design property of $M$ (Definition~\ref{defWellBehavedMatrixNewPaper}), 
\[|R|\leq t^2k.\]

    For every $b\in\nbcube{R}$, %
    let $\Dcal_b$ be the distribution obtained by conditioning $\Dcal$ to the event $\set{u\restrict{R} = b}$.

Note that, for every event $A$, we have that
\begin{align*}
\prob{x(u)\in\Dcal}{x(u)\in A}
&= \sum_{b\in\nbcube{R}}\prob{x(u)\in\Dcal}{u\restrict{R} = b}\prob{x(u)\in\Dcal}{x(u)\in A|u\restrict{R} = b} \\
&= \sum_{b\in\nbcube{R}}\prob{x(u)\in\Dcal}{u\restrict{R} = b}\prob{x(u)\in\Dcal_b}{x(u)\in A},
\end{align*}
thus, $\Dcal = (\Dcal_0)_S$ is a convex combination of the distributions $\setbar{\Dcal_b}{b\in\nbcube{R}}$.

Let us now prove that, for every $b\in\nbcube{R}$, the distribution $\Dcal_b$ is a product distribution on $x\in\bcube{S}$.
Note that, for every $i\in S$, we have
that 
\begin{align*}
x_i = 1
&\iff M[i]\cdot u = 0 \\
&\iff M[i]\restrict{\supp(M[i])\drop R}\cdot u\restrict{\supp(M[i])\drop R} = -M[i]\restrict{R}\cdot u\restrict{R}\\
&\iff \sum_{j\in \supp(M[i])\drop R} u_j = -\sum_{j\in\supp(M[i])\cap R}b_j\eqdef c_{b,i},
\end{align*}
where $c_{b,i}$ has value in the interval $[-t^2k, t^2k]$.\snote{It is also important that $c_{b,i} \equiv |R|\pmod{2}$?}\tnote{If it is not the case then the probabilities are zero. Below I mention that $N+v$ should be even.}
Let 
\[S_i\defeq \supp(M[i])\drop R.\]
Thus, for every $i,i'\in S$ with $i\neq i'$,
the variables $x_i$ and $x_{i'}$ of $\Dcal_b$ have disjoint sources of randomness
$\set{u_j}_{j\in S_i}$ and $\set{u_j}_{j\in S_{i'}}$.
Hence, the facts above imply that $\Dcal_b$ is a product distribution with each variable $x_i$ for $i\in S$ being a 
Bernoulli random variable with success probability
\[q_i\defeq
\prob{u}{\sum_{j\in S_i} u_j = c_{b,i}}.
\]
Therefore, the final step of the proof is to prove upper and lower bounds for $q_i$.
These bounds follow from standard calculations about the distribution of the height of a 1-dimensional random walk with $|S_i|$ steps.
For completeness, we prove the following lemma in the appendix.
\begin{lemma}
    \label{lemmaBoundsForRademacherRandomVariable}
    If the hypothesis of Lemma~\ref{lem:pseudo-ind-D0} holds,
    then there is a universal constant $c > 0$ such that
    \[\sqrt{\frac{2}{\pi s}}(1 - cs^{-1/3})
    \leq q_i
    \leq \sqrt{\frac{2}{\pi s}}(1 + cs^{-1/3}).\]
\end{lemma}
\begin{proof}
    See Appendix~\ref{appendixBoundsForRademacherRandomVariable}.
\end{proof}
This finishes the proof of Lemma~\ref{lem:pseudo-ind-D0}.

\end{proof}

The following lemma follows from Lemma~\ref{lem:pseudo-ind-D0} and Corollary~\ref{cor:robust-sf-main}.
\begin{corollary}
    \label{cor:sunflower-D0}
    Assume that $(s/k^5) \geq m^{\Omega(1)}$.
    Fix any $w\in\Naturals$ such that $w\leq (s/k^5)^{1/6}$.
    For any $r\geq s$,
    the distribution $\mathcal{D}_0$ satisfies the following property.
    For any $\ell\in [2w]$, any monotone DNF $\Fcal$ with at least $r^\ell$ terms of width $\ell$ contains a
    $(\mathcal{D}_0, m^{-3w})$-sunflower.  
\end{corollary}
\begin{proof}
    Let  
    \[
    \delta\defeq m^{-3w} \text{ and }
    t\defeq s^{1/3}/k^{3/2} \text{ and }
    q\defeq \Theta(1/\sqrt{s}).
    \]
    By Lemma~\ref{lem:pseudo-ind-D0} and the choice of $t$,
    the distribution $\mathcal{D}_0$ is $(t,1/t)$-pseudo-independent with marginal probability $q$.
    Note that $w$ satisfies $w\leq \sqrt{\frac{t}{(\log m)^2}}$ and
    \[w\log(1/\delta)
    \leq 3w^2\log m
    \leq o(t),
    \]
    so $\Dcal_0$ is $(t, \Omega(1/w))$-pseudo-independent and satisfies the hypothesis of Corollary~\ref{cor:robust-sf-main}.
    By Corollary~\ref{cor:robust-sf-main}
    (using $r'\defeq \Theta(w\log(1/\delta)/q)$ as the $r$ in the statement of this corollary
    and width $\ell\leq 2w$), we obtain the result above since $r \geq s\geq r'$.
\end{proof}

\subsubsection{The Monotone Boolean Circuit Lower Bound}

We now prove Lemma~\ref{lem:well-behaved}, the main lemma of this subsection.

\begin{proof}[Proof of Lemma~\ref{lem:well-behaved}]
It suffices to prove a lower bound for the monotone function $g$ on $m_1$ variables which is a restriction of $f.$

We apply Theorem~\ref{thm:sunflower-gen-cfmsy} with the distributions $\mathcal{D}_0,\mathcal{D}_1$ defined above. By Lemma~\ref{lem:alpha-D0} and Lemma~\ref{lem:alpha-D1}, we have
\[
\min\left\{\prob{x\sim \mathcal{D}_0}{f(x) = 0}, \prob{x\sim \mathcal{D}_1}{f(x) = 1}\right\}\geq \frac{1}{2}.
\]
By Lemma~\ref{lem:spreadness-D1} and Corollary~\ref{cor:sunflower-D0},
we see that Theorem~\ref{thm:sunflower-gen-cfmsy} is applicable with parameters 
\[Q \defeq \frac{m_1}{100n\log n} \text{ and } r \defeq s.\]
Note also that 
\[
8r = 8s \leq O\left(\frac{m_1}{n\cdot m^{\Omega(1)}}\right)\leq Q \leq rm_1.
\]
where we used the hypothesis $s\leq m_1/(n\cdot m^{\Omega(1)})$ of the lemma to get the second equality.

Applying Theorem~\ref{thm:sunflower-gen-cfmsy} now proves the claim.
\end{proof}

\subsection{Uniform Arithmetic Circuits for $(P_m)$}\label{sec:constructM}

In order to finish the proof of \Cref{thm:radical-sep-main},
we will need the following construction, whose proof we defer to the next section.

\begin{lemma}[Explicit construction of $(M_m)$]
    \label{lem:explicit-construction}
    There is a deterministic $\poly(m)$ time algorithm, which on input $1^m$, produces a matrix $M_m\in \{0,1\}^{n\times m}$ with the following properties.
    \begin{enumerate}
        \item $m = \Theta(n^{3/2}).$
        \item $M_m$ is $(C,s,k)$-well-behaved for a set $C$ of size
        $\Theta(m),$ $s = \Theta(\sqrt{n})$ and $k = O(1).$
    \end{enumerate}
\end{lemma}

Now we have all the ingredients to prove the main result of this section (Theorem~\ref{thm:radical-sep-main}).

\begin{proof}[Proof of Theorem~\ref{thm:radical-sep-main}]
We use the above sequence of matrices to define the polynomial sequence $(P_m)$ as in Section~\ref{sec:cfmsy}.
The upper bound follows from Lemma~\ref{lem:alg-ubd}.
The lower bound follows from Remark~\ref{obs:radicaltomonotone} and Lemma~\ref{lem:well-behaved}.    
\end{proof}

\section{Construction of Designs with Linear-Algebraic Expansion}
\label{sectExplicitConstructionRealCase}

The goal of this section is to prove the following theorem.

\begin{theorem}
\label{theoExplicitConstructionWellBehavedMatrix}
    There is a deterministic $\poly(q,s,t,d)$ time algorithm,
    which on input $(1^q, 1^s, 1^d, 1^t)$ for any $s,t,d\in\Naturals$ and prime number $q\in\Naturals$ such that
    $s\leq q$ and $d < q$ and $t\leq q^{d-1}$,
    outputs a matrix $G\in \{0,1\}^{n\times m}$ for
    \[n\defeq sq \text{ and } m\defeq m_1 + s \text{ and } m_1\defeq t((q-1)s+1)\]%
    such that
    $G$ is $(C,s,d)$-well-behaved matrix (Definition~\ref{defWellBehavedMatrixNewPaper}) for a set $C$ of size $m_1$.
\end{theorem}

Let us first see how the theorem yields \Cref{lem:explicit-construction}
from the previous section.
\begin{proof}[Proof of \Cref{lem:explicit-construction}]
    This lemma is a special case of the construction Theorem~\ref{theoExplicitConstructionWellBehavedMatrix} 
    using 
    parameters $k\defeq 2$ and
    \[
    q\defeq cn^{1/2}\text{ prime for $c > 2$, and }
    s\defeq 2\floor{n^{1/2}},\text{ and }  %
    t\defeq \frac{m - s}{(q-1)s + 1}\leq q = q^{k-1},  %
    \]
    and $n\defeq qs$.
\end{proof}

\noindent
We now prove \Cref{theoExplicitConstructionWellBehavedMatrix}.
Let $q, s, d, t\in\Naturals$ as in the statement of this theorem.
Consider the following notation.
\begin{definition}
\label{defNotationForExplicitConst}
For every $a\in [q], b\in [s]$:
\begin{itemize}
    \item let $e_a\in\Reals^q$ be the vector defined by $(e_a)_i\defeq 1_{\set{i=a}}$ for every $i\in[q]$;
    \item let $e_b\in\Reals^s$ be the vector defined by $(e_b)_i\defeq 1_{\set{i=b}}$ for every $i\in[s]$;
    \item let $\delta_{a,b}\in\rmatrixset{q}{s}$ be the matrix $\delta_{a,b}\defeq e_a\transpose{e_b}$;
    \item let $1_q\in\Reals^q$ be the vector $1_q\defeq \sum_{i\in [q]}e_i$;
    \item let $1_s\in\Reals^s$ be the vector $1_s\defeq \sum_{i\in [s]}e_i$;
    \item let $C_b\defeq 1_q\transpose{e_b}\in\rmatrixset{q}{s}$, which is a matrix consisting of the $b$-th column being the all-ones vector;
        \qedhere
\end{itemize}
\end{definition}
Let us start by defining a collection of matrices in $\bset^{q\times s}$ that
we will use to define the columns (vectors in $\bcube{qs}$) of the matrix $G\in\rmatrixset{n}{m}$ outputted by our algorithm.
For any function $f\ffrom [s]\fto \Fbb_q$, let
\[M_f\defeq \sum_{i\in[s]} e_{f(i)}\transpose{e_i}\in \Reals^{q\times s}\]
be the matrix representing the graph of the polynomial $f$.
Let $V\subseteq\Reals^{q\times s}$ be the following vector space
\[V\defeq \linspan\setbar{M_f}{f\ffrom [s]\fto \Fbb_q},\]
and let $\Ccal\subseteq\Reals^{q\times s}$ be the following collection of matrices
\[\Ccal\defeq \setbar{M_f}{\text{$f\ffrom [s]\fto \Fbb_q$ is a univariate
polynomial of degree at most $d$}}.\]
Let us first prove that $\Ccal$ satisfies a version of the design property
of Definition~\ref{defWellBehavedMatrixNewPaper}.
\begin{lemma}
    \label{lemmaDesignPropertyOfCcal}
    For every $A,B\in\Ccal$, if $A\neq B$, then
    \[|\supp(A)\cap\supp(B)|\leq d\]
\end{lemma}
\begin{proof}
    Let $f,g\in\Fbb_q[x]$ be univariate polynomials of degree at most $d$ such that
    \[A = M_f\text{ and } B = M_g.\]
    Note that if $(a,b)\in\supp(A)\cap\supp(B)\subseteq [q]\times [s]$, then
    \[f(b) = a = g(b).\]
    As $f$ and $g$ are distinct polynomials of degree at most $d$, they can intersect in at most $d$ points,
    which implies the conclusion of this lemma.
\end{proof}

Now our goal will be to develop tools that will help us to analyse the ``linear-algebraic expansion''
for the matrices in the collection $\Ccal$.
For every $a\in [q]$, $b\in [s]$ and $r\in\Fbb_q$,
let ${l_{a,b,r}\ffrom [s]\fto \Fbb_q}$ be the linear function defined by
\[l_{a,b,r}(i)\defeq (i - b)r + a\]
for every $i\in [s]$. Let 
\[M_{a,b,r}\defeq M_{l_{a,b,r}}.\]
Let us now define
\[
    \Acal\defeq\setbar{M_l}{\text{$l\ffrom [s]\fto \Fbb_q$ is a polynomial of degree at most $1$}}
    \text{ and }
    \Bcal\defeq\setbar{C_b}{b\in [s]}.
\]
Consider the following technical lemmas:
\begin{lemma}
    \label{lemmaDimensionOfVInExplicitConst}
    The vector space $V$ has dimension $d_V\defeq \dim(V) = (q-1)s + 1$.
\end{lemma}

\begin{proof}
    We first claim that
    \begin{equation}
        \label{eqCharacterizationOfSpaceOfMatrices}
        V = \setbar{M\in\rmatrixset{q}{s}}{\exists c\in\Reals\sttext \transpose{1_q}Me_i = c\; \forall i\in [s]}\eqdef W,
    \end{equation}
    that is, $V$ is the set of all matrices whose the sums of the entries of each of its columns are the same.
    As $V$ is the span of matrices $M_f$ and the sum of all the entries of each column of a matrix $M_f$ is 1, 
    we obtain that $V\subseteq W$.
    In order to prove that $W\subseteq V$,
    one can argue
    \bnote{should you not sketch the argument? or is that
    enough?} 
    \tnote{I sketched an argument below.}
    that every matrix $M\in W$ can be written as a linear combination of matrices $M_f$
    via the following arguments (induction on the number of nonzero entries of $M$):
    \begin{enumerate}
        \item If $M = 0$, then $M\in V$.
        \item Otherwise, if the column sum of $M$ is zero, then there is a column $i$ with two different entries $u, v$ with opposite nonzero value $a$.
        Define
        \[M' \defeq M - aM_f + M_g,\]
        where $f, g$ are the functions with $f(i) \defeq u$ and $g(i) \defeq v$ and with $f(j) = g(j)$ for a fixed value.
        Note that $M'\in W$ and has two less nonzero entries than $M$.
        \item If the column sum of $M$ is nonzero, then choose a nonzero element $j_i$ for every column $i$ of $M$ and then let
        \[M' \defeq M - aM_f\]
        where $a\defeq M_{j_1, 1}$ and $f(i) \defeq j_i$.
    \end{enumerate}

    Now we come back to the proof of the lemma.
    Let $\phi\ffrom \rmatrixset{q}{s}\fto \Reals^s$ be the following linear map:
    \[\phi(M) \defeq \sum_{i\in [s]} (\transpose{1_q}Me_i)e_i.\]
    By the characterization in Equation~\ref{eqCharacterizationOfSpaceOfMatrices}, we obtain
    \[\ker(\phi) = \setbar{M\in\rmatrixset{q}{s}}{\transpose{1_q}Me_i = 0\; \forall i\in [s]}\subseteq V,\]
    and $\phi(V) \isom \Reals$.
    By the rank-nullity theorem,
    \[qs = \dim(\rmatrixset{q}{s})
    = \dim(\ker(\phi)) + \dim(\Image(\phi))
    = \dim(\ker(\phi)) + s,
    \]
    so $\dim(\ker(\phi)) = (q-1)s$.
    Let $\psi\defeq \phi\restrict{V}\ffrom V\fto \Reals^s$.
    Thus, $\ker(\psi) = \ker(\phi)$ and $\Image(\psi) = \phi(V) \isom \Reals$,
    and, by the rank-nullity theorem,
    \[\dim(V) = \dim(\ker(\psi)) + \dim(\Image(\psi)) = (q-1)s + 1.\]
\end{proof}

\begin{lemma}
    \label{lemmaCollectionAcalAndBcalSpanEverything}
    The set 
    $\Acal\cup\Bcal$
    spans $\rmatrixset{q}{s}$.
\end{lemma}
\begin{proof}
    Let us prove that the set $\Dcal\defeq\Acal\cup\Bcal$ is a spanning set of $\rmatrixset{q}{s}$.
    For every $a\in [q], b\in [s]$, note that
    \[C_b + \sum_{r\in\Fbb_q}M_{a,b,r}
    = q\delta_{a,b} + \sum_{b'\in [s]} C_{b'},\]%
    since the support of the matrices $C_b$
    and $\set{M_{a,b,r}}_{r\in [q]}$ only pairwise
    intersect entry $(a,b)$, %
    and every other entry in $[q]\times [s]$ is in the support of exactly one of these matrices.
    Hence, $\delta_{a,b}$ is in $\linspan(\Dcal)$,
    which implies that $\Dcal$ is a spanning set of $\rmatrixset{q}{s}$.
\end{proof}

\begin{lemma}
    \label{lemmaIndMatrixOfAffinePolysSpansV}
    The set  $\Acal$ spans $V$.
\end{lemma}
\begin{proof}
    Let us prove that $\Acal$ is a spanning set of $V$.
    First note that if $M\in V$,
    then there is $c\in\Reals$ such that the sum of the entries in any
    column of $M$ is equal to $c$ (as this is true for any matrix $M_f$ in
    the definition of $V$; see the proof of (\ref{eqCharacterizationOfSpaceOfMatrices}) above).\bnote{The proof of the claim in this sentence
        is hidden in the
        proof of Lemma 4.4 above. Perhaps the claim should be proved
        separately in a claim that you cite?}
    Thus, for every $M\in V\cap\linspan(\Bcal)$,
    there is $\alpha\in\Reals^s$ such that 
    \[M = \sum_{i\in [s]} \alpha_i C_i,\]
    and there is $c\in\Reals$ such that $\transpose{1_q}(Me_i) = c$ for every $i\in [s]$.
    As the matrices $C_i$'s are column matrices with disjoint support, we get that,
    for every $i\in [s]$,
    \[c = \transpose{1_q}(Me_i)
    = \alpha_i\transpose{1_q}{(C_ie_i)} = \alpha_i q,\]
    which implies that
    \begin{align*}
        M
        &= \sum_{i\in [s]} \alpha_i C_i \\
        &= \frac{c}{q}\sum_{i\in [s]} C_i \\
        &= \frac{c}{q}\sum_{a\in [q]} M_{a,1,0}\in\linspan(\Acal).
    \end{align*}
    Therefore, $V\cap\linspan(\Bcal) \subseteq \linspan(\Acal)$,
    which implies that $\Acal$ spans $V$,
    since $\Acal\subseteq V$ and Lemma~\ref{lemmaCollectionAcalAndBcalSpanEverything}.
\end{proof}

\begin{lemma}
    \label{lemmaIndMatrixOfFunctionPlusAffinePolysSpansV}
    For every function $f\ffrom [s]\fto \Fbb_q$,
    the set 
    \[\Acal_f\defeq\setbar{M_{f+l}}{\text{$l\ffrom [s]\fto \Fbb_q$ affine polynomial}}\]
    spans $V$.
\end{lemma}
\begin{proof}
    Let 
    \[\text{$f\ffrom [s]\fto \Fbb_q$ be a function and $g\defeq -f$.}\]
    Let $\pi_f\ffrom V\to\rmatrixset{q}{s}$ be the following map:
    for every $M\in V$ and every $i\in [s]$,
    let $M_i$ be the $i$-th column of $M$ and
    define the $i$-th column $P_i$ of $\pi_f(M)$ to be
    \[
        P_i\defeq \sum_{j\in [q]} (M_i)_{j-f(i)}e_j.
    \]
    You can interpret $P_i$ as a cyclic shift of the column $M_i$ using the value of $f(i)$.
    Note that $\pi_f$ is a bijective linear map with inverse $\pi_g$,
    and, for any linear function $l$,
    \[\pi_f(M_l) = M_{f+l}.\]
    Hence, 
    \[V = \text{$\pi_f(V)$, and
      $\Acal_f = \pi_f(\Acal)$,}\]
    which implies that $\Acal_f$ spans $V$,
    since, by Lemma~\ref{lemmaIndMatrixOfAffinePolysSpansV}, $\Acal$ spans $V$.
\end{proof}

We use the lemma above to construct the matrix $G\in\bcube{n\times m}$ and analyse its linear-algebraic expansion.

For every $\alpha\in \Fbb_q^{d-1}$, let 
\[F_\alpha(x)\defeq \sum_{i=1}^{d-1} \alpha_ix^{i+1}.\]
Using Lemma~\ref{lemmaIndMatrixOfFunctionPlusAffinePolysSpansV}, fix a basis $\Bcal_\alpha\subseteq \Acal_{F_\alpha}$ for $V$.
By Lemma~\ref{lemmaDimensionOfVInExplicitConst}, we have that 
\[|\Bcal_\alpha| = \dim(V) = (q - 1)s + 1 = d_V.\]
As $d < q$, two polynomials of degree at most $d$ do not correspond to the same function;
thus, for every $\alpha,\beta\in\Fbb_q^{d-1}$ distinct, we have 
\[
    \Acal_{F_\alpha}\cap \Acal_{F_\beta} = \emptyset
    \text{ and }
    \Bcal_\alpha\cap\Bcal_\beta = \emptyset,
\]
since $\mathsf{deg}(F_\alpha), \mathsf{def}(F_\beta) \geq 2$.

Let $\Scal_t\subseteq \Fbb_q^{d-1}$ be an arbitrary fixed set of $t$ elements of $\Fbb_q^{d-1}$
(e.g., the first $t$ elements according to the lexicographic order).
Define
\[\Ccal'\defeq\cup_{\alpha\in \Scal_t}\Bcal_\alpha.\]
Hence, $|\Ccal'| = td_V$ and,
as $\Ccal'\subseteq \Ccal$, Lemma~\ref{lemmaDesignPropertyOfCcal} implies that $\Ccal'$ also satisfies a design property.
Now we show that $\Ccal'$ satisfies a linear-algebraic expansion.

\begin{lemma}
    \label{lemmaLinearAlgebraicExpandingPropOfCcal}
    For every subspace $U\subseteq V$, the probability that a uniformly random matrix of $\Ccal'$ lie in $U$ is at most $\dim(U)/\dim(V)$.
\end{lemma}
\begin{proof}
    Consider the following random process to generate a uniform random matrix $M$ of $\Ccal'$:
    \begin{enumerate}
        \item Choose $\alpha\in\Scal_t$ uniformly at random;
        \item Let $M$ be a uniformly random matrix in $\Bcal_\alpha$.
    \end{enumerate}
    It is not hard to prove that $M$ is a uniform random matrix of $\Ccal'$ (as $\Bcal_\alpha$'s are disjoint),
    and, if we condition on $\alpha\in\Scal_t$, then $M$ is a uniform random matrix of $\Bcal_\alpha$.
    For every $a\in\Scal_t$, we have that
    \[|\Bcal_a\drop U| \geq \dim(V) - \dim(U)\]
    as $\Bcal_a$ is a basis of $V$.
    Hence,
    \[\Pr_{M,\alpha}[M\not\in U|\alpha = a]
      \geq \frac{\dim(V)-\dim(U)}{|\Bcal_a|}
      = 1 - \frac{\dim(U)}{\dim(V)}.
    \]
    Therefore, by a conditioning argument over the component $\alpha$ of the random process generating $M$, we get that
    \[\Pr_{M,\alpha}[M\not\in U]
      \geq 1 - \frac{\dim(U)}{\dim(V)},
    \]
    which proves the lemma.
\end{proof}

Therefore, we can construct the matrix $G\in\bcube{n\times m}$ as follows.
Recall that 
\[n = qs\text{ and }m = m_1 + s\text{ and }m_1 = td_V.\]
Fix total orders for the elements of $\Scal_t$ and the elements of each $\Bcal_\alpha$.
For every $i\in [t]$ and $j\in [d_V]$,
let $\alpha\in\Scal_t$ be the $i$-th element of $\Scal_t$
and $M\in\Bcal_\alpha$ be the $j$-th element of $\Bcal_\alpha$.
We define the $((i-1)t + j)$-th column $G_{i,j}$ of $G$ to be
\[G_{i,j}\defeq \mathtt{vect}(M)\in\bcube{qs},\]
where $\mathtt{vect}\ffrom\rmatrixset{q}{s}\fto\Reals^{qs}$ is a
fixed isomorphism between matrices $M\in\bcube{q\times s}$ and a vector $v\in\bcube{qs}$
obtained by rearranging the $qs$ entries of $M$ as entries of $v$ in a fixed arbitrary way.
For every $j\in [s]$, we define the $(m_1 + i)$-th column $G_{0,j}$ of $G$ to be
\[G_{0,j}\defeq \mathtt{vect}(C_j),\]
where $C_j$ as defined by Definition~\ref{defNotationForExplicitConst}.
It is not hard to show that $G$ can be constructed in time $\poly(q, s, t,
d)$ by a deterministic algorithm, as the sets $B_\alpha$ can be obtained by findinig a basis of the set $\Acal_\alpha$.
\bnote{Maybe say how to compute the $B_\alpha$'s.}
\tnote{Added a brief explanation}

For
\[C\defeq \setbar{(i-1)t + j}{i\in [t], j\in [d_V]}\subseteq [m],\]
let us prove that the matrix $G\in\bcube{n\times m}$ defined above is $(C, s, d)$-well behaved:
\begin{itemize}
    \item The vector space $V'\defeq \linspan\setbar{G_{i,j}}{i\in [t], j\in [d_V]}$ of column of $G$ indexed by $C$
        is isomorphic to
        the vector space $V$ spanned by the matrices in $\Ccal'$ using the isomorphism $\mathtt{vect}$ above.
    \item All the matrices $M\in\Ccal'$ and $C_j$'s have support of size $s$,
        and the columns of $G$ are rearrangements of the entries of these matrices in vector form.
        Thus, all columns of $G$ have support size $s$.
    \item By Lemma~\ref{lemmaDesignPropertyOfCcal},
        all pairs of distinct matrices $M, N$ in $\Ccal'$ have
        \[|\supp(M)\cap\supp(N)|\leq d.\]
        This also holds for all pairs of columns of $G$ in $C$ as they are vector corresponding to matrices in $\Ccal'$.
    \item For any subspace $U'\subseteq V'$,
        the probability of a column of $G$, chosen uniformly at random from the columns indexed by $C$, be inside $U'$
        is the same as
        the probability of an element of $\Ccal'$, chosen uniformly at random, be inside a subspace $U$ of $V$ isomorphic to $U'$.
        By Lemma~\ref{lemmaLinearAlgebraicExpandingPropOfCcal},
        the latter probability is at most $\dim(U')/\dim(V')$.
\end{itemize}

\bibliographystyle{alphaurl}
\bibliography{refs.bib} %

\appendix

\section{Properties of $\mathcal{D}_1$}
\label{appendixPropertiesOfD1}

\begin{proof}[Proof of Lemma~\ref{lem:spreadness-D1}]
Let $T\defeq\set{u_1,\dotsc, u_t}\subseteq [m_1]$.
By Definition~\ref{defHardDistributions},
$x\sim\Dcal_1$ is the indicator vector of a random set $I$ obtained from sampling $\lambda$ elements $v_1,\dotsc, v_\lambda$ of $[m_1]$ with replacement.
Thus
\[
\prob{x\sim\Dcal_1}{\land_{i\in T}x_i = 1}
\leq \sum_{S\in \binom{[\lambda]}{t}}
     \prob{I}{\cap_{i\in S} \set{v_i = u_i}} 
=  \sum_{S\in \binom{[\lambda]}{t}}
     m^{-t}
\leq \lambda^tm^{-t}
= (m/\lambda)^{-t}.
\]
\end{proof}

Our proof of Lemma~\ref{lem:alpha-D1} is essentially the same proof as for a similar lemma in~\cite{CFMSY25}, but using the linear-algebraic expansion property of the matrix.
\begin{proof}[Proof of Lemma~\ref{lem:alpha-D1}]
By the definition of $g$, we have that, for every $S\subseteq [m_1]$,
\[g(1_S) = 1\iff V = \linspan\setbar{M[i]}{i\in S}.\]
By Definition~\ref{defHardDistributions},
$x\sim\Dcal_1$ is the indicator vector of a random set $I$ obtained from sampling $\lambda$ elements $v_1,\dotsc, v_\lambda$ of $[m_1]$ with replacement.
For every $i\in [\lambda]$, let 
\[C_i\defeq M[v_i].\]
Let $N\defeq \dim(V)$.
For each $d \leq N$,
let $X_d$ denote the smallest $r\in [\lambda]$ such that
$C_1,\ldots, C_r$ span a subspace of dimension $d$ (if such an $r$ exists).
For $X_0 \defeq 0$,
we have that 
\[X_N = \sum_{d=0}^{N-1} X_{d+1}-X_{d},\]
which implies that
\[\Exp[X_N] = \sum_{d=0}^{N-1} \Exp[X_{d+1}-X_{d}].\]
For every $d\in [N]$ and $r\in [\lambda]$,
if $X_d = r$, then 
\[U\defeq \linspan\set{C_1,\dotsc, C_r}\]
has dimension $d$.
Thus,
by the linear-algebraic expansion property for $(C, s, k)$-well-behaved matrices (Definition~\ref{defWellBehavedMatrixNewPaper}),
we have that, for every $i > r$,
\[\prob{x}{C_i\in U}
\leq \frac{\dim(U)}{\dim(V)}
= \frac{d}{N}
\eqdef p_d.\]
It follows that $X_{d+1}-X_d$ (conditioned on $v_1,\ldots, v_r$ spanning a space of dimension $d$)
has a geometric distribution with success probability at least 
$1-p_d = (N-d)/N$ and thus has expectation at most $N/(N-d)$. 
Hence,
\[
\Exp[X_N] \leq \sum_{d=0}^{N-1} \frac{N}{N-d} \leq 2N\log N. 
\]
By Markov's inequality,
\[
\prob{x}{X_N > \lambda}
\leq \prob{x}{X_N > 100n\log n}
\leq \prob{x}{X_N > 100N\log N}
\leq \prob{x}{X_N > 50\Exp[X_N]}
\leq 1/50.
\]
This concludes the proof of the lemma, as
\[g(x) = 1\iff X_N\leq \lambda.\]
\end{proof}

\section{Proof of Lemma~\ref{lemmaBoundsForRademacherRandomVariable}}
\label{appendixBoundsForRademacherRandomVariable}   

Let $v\in [0,|R|]$ and $N\geq v$ such that $N + v$ is even.
Let $R_N\sim\mathrm{Rademacher}(N)$ be a sum of $N$ independent Rademacher random variables.
A simple counting argument shows that
\[\prob{}{R_N = v} = \binom{N}{(N+v)/2}2^{-N}.\]
Furthermore, for
$N = 2l + b$ for $l\in\Naturals$ and $b\in\set{0,1}$
and
$w\defeq (v+b)/2$, we obtain
\begin{align*}
    \frac{\binom{N}{(N+v)/2}}{\binom{N}{\ceiling{N/2}}}
    &= \frac{\ceiling{N/2}!\floor{N/2}!}
            {((N+v)/2)!((N-v)/2)!}
     = \frac{(l + b)!l!}
            {(l + (b+v)/2)!(l + b - (v+b)/2)!} \\
    &= \frac{\prod_{i = 0}^{w-1} l + b - i}
            {\prod_{i = 1}^{w} l + i}
     = \prod_{i = 1}^{w} \left(1 - \frac{2i - b - 1}{l + i}\right). \\
\end{align*}
Using that $\exp(-x/(1-x))\leq 1-x\leq \exp(-x)$ %
for every $x\leq 1$,
we obtain that
\begin{align*}
     \prod_{i = 1}^{w} \left(1 - \frac{2i - b - 1}{l + i}\right)
     &\leq \exp\left(-\sum_{i=1}^w \frac{2i - b - 1}{l + i}\right) \\
     &\leq \exp\left(-\frac{1}{l+w}(1-b+2w-b-1)w/2\right) \\
     &= \exp\left(-\frac{(w-b)w}{l+w}\right),
\end{align*}
and
\begin{align*}
     \prod_{i = 1}^{w} \left(1 - \frac{2i - b - 1}{l + i}\right)
     &\geq \exp\left(-\sum_{i=1}^w \frac{2i - b - 1}{l - i + b + 1}\right) \\
     &\geq \exp\left(-\frac{1}{l-w}(1-b + 2w-b-1)w/2\right)
      = \exp\left(-\frac{(w-b)w}{l-w}\right) \\
     &= \exp\left(-\frac{(w-b)w}{l+w}\left(1+\frac{2w}{l-w}\right)\right) \\
     &= \exp\left(-\frac{(w-b)w}{l+w}\right)
        \exp\left(-\frac{2(w-b)w^2}{(l+w)(l-w)}\right).
\end{align*}
Hence,
\[\binom{N}{(N+v)/2}
\in \binom{N}{\floor{N/2}}
    \cdot \exp\left(-\frac{(w-b)w}{l+w}\right)
    \cdot \left[\exp\left(-\frac{2(w-b)w^2}{(l+w)(l-w)}\right), 1\right].
\]
In the following, consider 
\[c_1,\dotsc, c_{10}\in\Reals_{\geq 0}\]
to be positive constants implicitly defined during our calculations.
It is well known that 
\[n! \in \sqrt{2\pi n}(n/e)^n\cdot [1-c_1/n, 1+c_1/n]\]
for sufficiently large $n$.\tnote{reference?}
So, we can prove that
\[\binom{N}{\floor{N/2}} \in \sqrt{\frac{2}{\pi N}}\cdot 2^N\cdot e^{c_2/N}\cdot [e^{-c_3/N}, 1].\]
Recall that, for 
\[N\defeq |S_i| = |\supp(M[i])\drop R|\in [s-t^2k, s], \text{ and }
v\defeq |c_{b,i}|\in [0, t^2k],
\]
we have
\[q_i = \prob{}{R_N = v}.\]
By our assumption, %
we have
$v^3\leq |R|^3 \leq (t^2k)^3\leq s$, so 
\[N
\in s\cdot \left[1-\frac{|R|}{s}, 1\right]
\subseteq s\cdot \left[1-\frac{1}{s^{2/3}}, 1\right]
\subseteq s\cdot [e^{-2s^{-2/3}}, 1],
\]
and $1/\sqrt{N}\in (1/\sqrt{s})e^{2s^{-2/3}}\cdot [e^{-2s^{-2/3}}, 1]$ and
\begin{align*}
    \prob{}{R_{N} = v}
    =&\quad   \binom{N}{(N+v)/2}\cdot 2^{-N} \\
    \in&\quad \sqrt{\frac{2}{\pi N}}\cdot e^{c_2/N}
              \cdot  e^{-(w-b)w/(l+w)}
              \cdot [e^{-c_3/N}e^{-2w^3/((l+w)(l-w))},
                     1] \\
    \subseteq&\quad \sqrt{\frac{2}{\pi s}}e^{2s^{-2/3}}\cdot e^{c_4(1+v^2)/s}
                    \cdot \left[e^{-2s^{-2/3}}e^{-(c_5/s)(1 + v^3/s)},
                                1\right].
\end{align*}
Moreover, we have $v^2/s\leq 1/s^{1/3}$, so
\begin{align*}
    \prob{}{R_{N} = v}
    \in      &\quad \sqrt{\frac{2}{\pi s}}e^{2s^{-2/3}}\cdot e^{c_6s^{-1/3}}
                    \cdot \left[e^{-2s^{-2/3}}e^{c_7s^{-1/3}}e^{-c_8s^{-1}},
                                1\right]. \\
    \subseteq&\quad \sqrt{\frac{2}{\pi s}}
                    \cdot \left[e^{-c_{9}s^{-1/3}},
                                e^{c_{10}s^{-1/3}}\right].
\end{align*}
Therefore,
\[\sqrt{\frac{2}{\pi s}}(1 - c_9s^{-1/3})
\leq q_i
\leq \sqrt{\frac{2}{\pi s}}(1 + 2c_{10}s^{-1/3}),
\]
and we define $c\defeq \max\set{c_9, 2c_{10}}$.

\end{document}